\documentclass[twocolumn,aps,amsmath,amssymb]{revtex4-2}
\usepackage[T1]{fontenc}
\usepackage{graphicx}
\usepackage{braket}
\usepackage[version=4]{mhchem}
\newcommand{\legendfont}[1]{{\fontfamily{InriaSans-LF}\fontseries{l}\selectfont #1}}
\usepackage[colorlinks, citecolor=blue, urlcolor=blue, linkcolor=red]{hyperref}
\usepackage{footnote}
\usepackage[referable]{threeparttablex}
\usepackage{multirow}
\usepackage{fancyhdr}

\begin{document}
	
	\title{Probability distribution analysis of the cascaded variational quantum eigensolver} 
	\author{Yi-Hua Lai}
	\affiliation{National Research Council Research Associate at the U.S. Naval Research Laboratory, Washington, DC 20375, USA}
	\author{John P.T. Stenger}
	\author{Gloria Bazargan}
	\author{Igor V. Schweigert}
	\author{Daniel Gunlycke}
	\email{lennart.d.gunlycke.civ@us.navy.mil}
	\affiliation{U.S. Naval Research Laboratory, Washington, DC 20375, USA}
	
	\begin{abstract}
		The cascaded variational quantum eigensolver (CVQE) circumvents the need for iterative communication between the quantum and classical processing units that is necessary in the conventional VQE algorithm. While CVQE offers complete freedom to choose the guiding state as input, not all guiding states provide both accurate solutions and efficient resource usage. Our work presents a process based on trapezoidal-state preparation for selecting guiding states that yield accurate many-electron ground-state solutions with minimal resource consumption. By analyzing state probability distributions at different stages of CVQE calculations for the collision complex mediating the $\ce{H_2 + H_2^+\rightarrow H_3^+ + H}$ bimolecular reaction, we identify stage-specific errors, trace their impact on final classical optimization, and determine the optimal guiding-state parameters for given resource constraints. Our findings uncover a critical trade-off on noisy intermediate-scale quantum (NISQ) devices, demonstrating that shallower diabatic circuits provide the best compromise among optimization efficiency, state quality, sampling cost, and signal-to-noise ratio in our system.
	\end{abstract}
	
	\maketitle
	
	\thispagestyle{fancy}
	
	\section{Introduction}\label{s.1}
	Quantum computing has the potential to revolutionize quantum chemistry because the exponentially large Hilbert space that describes many-particle systems can be represented by a linear number of qubits. Finding ground states using efficient quantum simulation algorithms is thus a promising research area. One approach to find the ground state is to implement quantum evolution operators directly, as in quantum phase estimation~\cite{kitaev1995quantum} and adiabatic state preparation~\cite{van2001powerful}. However, because of quantum decoherence, these deep-circuit approaches are not suitable for noisy intermediate-scale quantum (NISQ) computing~\cite{preskill2018quantum}. One workaround is the variational quantum eigensolver (VQE) algorithm~\cite{peruzzo2014variational, mcclean2016theory, o2016scalable, kandala2017hardware, wang2019accelerated, google2020hartree, gonthier2022measurements, mcardle2020quantum, cerezo2021variational, head2020quantum}. VQE requires fewer gates and is therefore less affected by quantum noise. The downside of this approach is that it entails iterative cycling between the QPU and CPU, necessitating a large number of quantum circuit executions. In contrast, by separating the optimization and sampling processes, the cascaded variational quantum eigensolver (CVQE) algorithm~\cite{gunlycke2024cascaded} does not require this QPU-CPU cycle. In CVQE, the trial state is dependent on a guiding state prepared and sampled on the QPU once to specify the Ansatz over which the trial state is subsequently optimized on the CPU. Because any Ansatz only covers a subset of the solution space, it is advantageous to choose a guiding state that produces an Ansatz close to the ground state. This can be achieved by discretizing the evolution operator using trapezoidal-state preparation~\cite{gunlycke2025guided, stenger2026hybrid}.
	
	In this paper, we apply the CVQE algorithm to investigate how trapezoidal-state preparation parameters affect probability distributions. Specifically, we compare the (1) trapezoidal-state, (2) guiding-state, (3) sampled, (4) optimized, and (5) ground-state probability distributions in order to optimize the state preparation parameters. These approaches are benchmarked on the prototypical bimolecular reaction
	\begin{equation}\label{reaction}
		\ce{H_2 + H_2^+\rightarrow H_3^+ + H}
	\end{equation} 
	Reaction \eqref{reaction} is responsible for the conversion of $\ce{H_2^+}$ to $\ce{H_3^+}$ in interstellar space~\cite{herbst1973formation,watson1973rate,prasad1974jovian} and serves as a key pathway to the production of more complex molecules~\cite{oka2006interstellar}. We present a stage-by-stage analysis of CVQE probability distributions to isolate state errors. Furthermore, since not all state errors compromise final optimization outcomes, this diagnostic capability allows us to design hardware-efficient circuits that target only the most critical errors.
	
	Section~\ref{s.2} reviews the CVQE and trapezoidal-state preparation methods. In Sec.~\ref{s.3}, we present a demonstration of our approach for reaction \eqref{reaction} in three different regimes for the trapezoidal-state preparation. Based on the results, Sec.~\ref{s.4} offers guidelines for choosing guiding-state preparation parameters.
	
	\section{Methods}\label{s.2}
	\subsection{CVQE}\label{s.2_1}
	CVQE is a quantum algorithm for finding the ground-state energy of a Hamiltonian, where a guiding state is prepared on a quantum computer and samples from the guiding state are used to generate a trial state that is optimized classically. The guiding state must be a reasonable approximation of the ground state of the Hamiltonian. To achieve this, we use a guided-sampling Ansatz (GSA) that is a crude approximation of adiabatic state preparation. A trial state is generated from the samples of the guiding state through a general operator $\hat \lambda$. The ground-state energy is then found by optimizing the trial state. The energy of a many-electron system described by the Hamiltonian $\hat{H}$ in the trial state $|\Psi(\theta)\rangle$ is the expectation value
	\begin{equation}\label{ExpE}
		E(\theta)=\frac{\langle\Psi(\theta)|\hat{H}|\Psi(\theta)\rangle}{\langle\Psi(\theta)|\Psi(\theta)\rangle},
	\end{equation}
	where $\theta$ is a collection of variational parameters. In quantum chemical calculations, one typically seeks the ground state $|\Psi_\text{g}\rangle$ of $\hat{H}$ and its associated energy $E_\text{g}$.  This energy can be estimated by the optimized energy $E^*=\min_{\theta}E(\theta)$, which, by the variational theorem, is an upper bound of $E_\text{g}$.
	
	In the CVQE~\cite{gunlycke2024cascaded} method, the trial state has the form 
	\begin{equation}\label{Ansatz}
		|\Psi(\theta)\rangle=e^{i\hat{\lambda}(n_{\mathcal M\mathcal{S}}, \theta)}|\Psi_0\rangle,
	\end{equation}
	where $\hat{\lambda}$ is an operator that depends on the samples $n_{\mathcal M\mathcal{S}}$ discussed below and the guiding state $|\Psi_0\rangle=\hat{U}|0\rangle$, where $|0\rangle$ denotes the tensor product with all the qubits in the zero state. The trial state in Eq.~\eqref{Ansatz} allows the optimization of the parameters $\theta$ to be performed completely on the CPU using the samples collected on the QPU.
	
	GSA is a class of Ansätze that uses system interactions to define the guiding state through $\hat{U}$ such that the Ansatz is in the neighborhood of $|\Psi_\text{g}\rangle$~\cite{gunlycke2025guided}. Without losing the expressivity of any Ansatz, we consider only a single-measurement basis in the Fock basis $\{|n\rangle\}$, where $n$ is a collection of number-operator eigenvalues, and drop $\mathcal M$ for brevity. The sample $n_\mathcal S$ is then the collection $(n_s)_{s\in\mathcal S}$ of outcomes $n_s$ over the shot set $\mathcal S$ obtained by measuring the guiding state $|\Psi_0\rangle$ with probability $|\langle n|\Psi_0\rangle|^2$. The total required sample size tends to be larger with the multi-basis approach, though the required sample for each basis may be smaller~\cite{stenger2026hybrid}. We also ignore multiplicity herein, focus on the outcome set $\mathcal N_\mathcal S=\{n_s\}_{s\in\mathcal S}$, and define the operator $\hat{\lambda}(n_\mathcal{S}, \theta)=\sum_n \lambda_n(\theta)|n\rangle\langle n|$ by the parametric equation
	\begin{equation}\label{Lambda_n}
		\lambda_n(\theta)=
		\begin{cases}
			-i\ln\frac{\theta_n}{\langle n|\Psi_0\rangle} & \text{for $n\in \mathcal N_\mathcal{S}$,}\\
			i\infty & \text{otherwise,}
		\end{cases} 
	\end{equation}
	where the parameters $\theta_n\in\mathbb{C}$ are elements of $\theta=(\theta_n)_{n\in\mathcal N_\mathcal{S}}$. With these choices, Eq.~\eqref{Ansatz} becomes
	\begin{equation}\label{Psi_sub}
		|\Psi(\theta)\rangle=\sum_{n\in \mathcal N_\mathcal{S}}\theta_n|n\rangle.
	\end{equation}
	As the cardinality of $\mathcal N_\mathcal{S}$, which is equal to or less than the number of shots, must increase at most polynomially with the system size in any quantum or classical algorithm, we can use the CPU to diagonalize the GSA Hamiltonian
	\begin{equation}\label{H_eff}
		\hat{H}_{\mathcal N_{\mathcal S}}=\sum_{n,n'\in \mathcal N_\mathcal{S}}h_{nn'} |n\rangle\langle n'|
	\end{equation}
	represented on the Fock subspace spanned by $\{|n_s\rangle\}_{s\in\mathcal S}$, where the coefficients $h_{nn'}=\langle n|\hat{H}|n'\rangle$.  The lowest eigenvalue of $\hat{H}_{\mathcal N_{\mathcal S}}$ is our ground-state estimate $E^*$, and the corresponding eigenstate $|\Psi^*\rangle=|\Psi(\theta^*)\rangle$ is the optimized state.
	
	\subsection{Trapezoidal- and guiding-state preparation}\label{s.2_2}
	As expanding system sizes exponentially dilute the ground state within the overall Fock space, preparing an initial state with high ground-state overlap provides critical leverage for quantum algorithms. One approach for preparing the guiding state close to the ideal ground state is through adiabatic state preparation~\cite{farhi2001quantum, van2001powerful, smelyanskiy2001simulations}. Let us introduce the parameter $\eta\in[0,1]$ and define the parameterized Hamiltonian
	\begin{equation}\label{AdiabaticH}
		\hat{H}(\eta)=(1-\eta)\hat{H}_0+\eta\hat{H}
	\end{equation}
	starting from a model Hamiltonian $\hat{H}_0$ at $\eta=0$ and ending at the system Hamiltonian $\hat{H}$ at $\eta=1$. For molecular systems, a natural choice is to use the Hartree-Fock (HF) Hamiltonian for $\hat{H}_0$ and the HF ground state $|\Phi_0\rangle=\hat U_0|0\rangle$ as our starting state at $\eta=0$.  The ideal guiding state is then given at $\eta=1$ by the unitary operator
	\begin{equation}\label{AdiabaticU}
		\hat{U}=\mathcal{T}\exp\left\{-i\beta K\int_{0}^{1}\hat{H}(\eta)d\eta\right\}\hat{U}_0,
	\end{equation}
	where $K$ is an integer, $\beta$ is the evolutionary step size, and $\mathcal{T}$ orders operators such that those with higher $\eta$ appear to the left. The discrete adiabatic theorem ensures that the distance between the guiding state and the ground state is arbitrarily small if $\beta$ is sufficiently small and the product $\beta K$ is sufficiently large.
	
	\begin{figure}
		\includegraphics[scale=0.4]{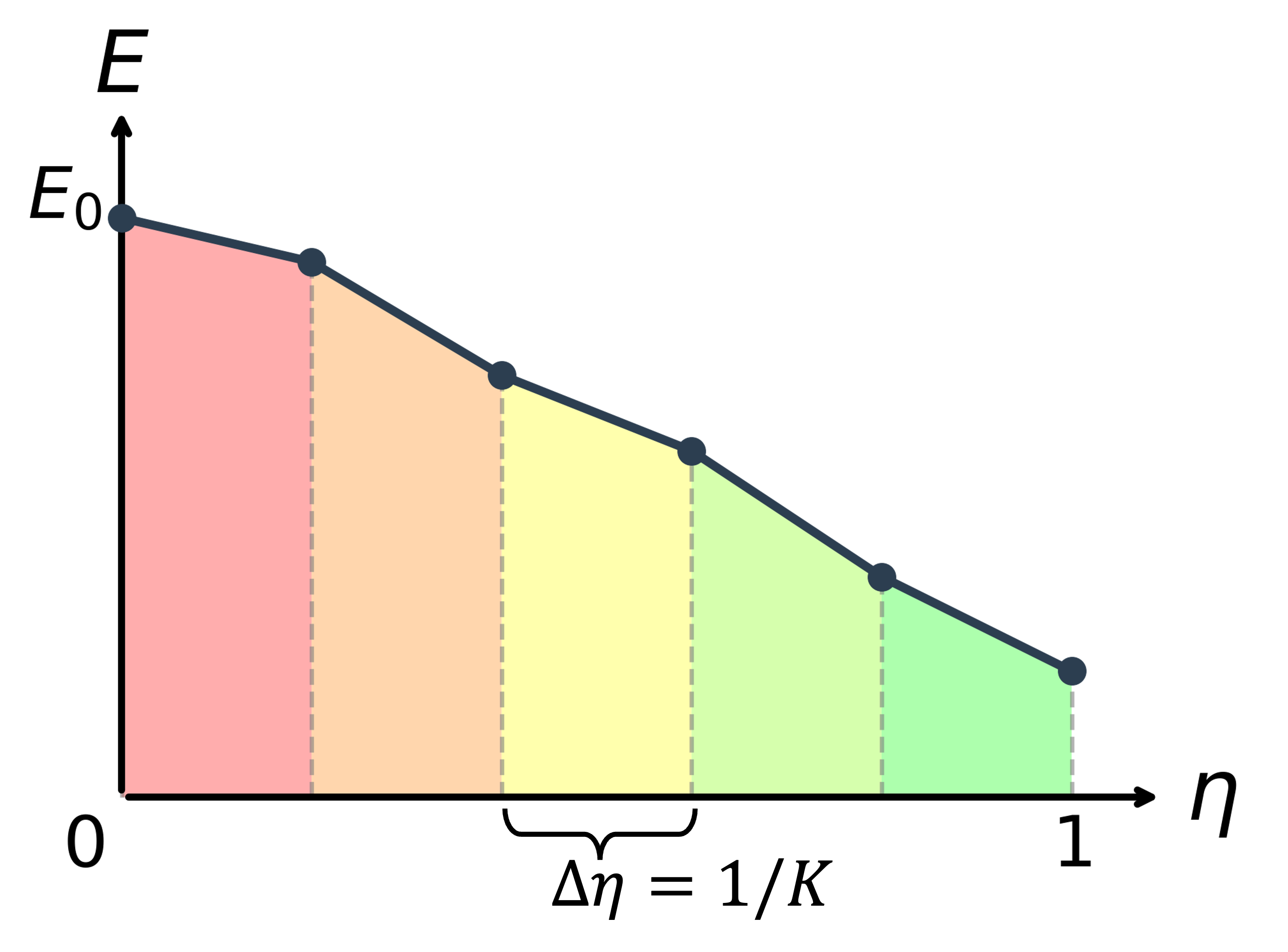}
		\caption{A schematic diagram of the energy $E(\eta)$ of the state evolving from the model ground state $|\Phi_0\rangle$ to the trapezoidal state $\hat{U}_{\beta, K}|\Phi_0\rangle$ as a function of the discretized parameter $\eta_k$ with $K=5$ steps of size $\Delta \eta=1/K$.}
		\label{f.1}
	\end{figure}
	
	In practice, we discretize $\eta$ and choose $K$ equal steps $\Delta \eta$ such that $\eta_k=k/K$ as shown in Fig.~\ref{f.1}. Applying the trapezoidal rule and the generalized Trotter formula~\cite{suzuki1976generalized, suzuki1977convergence, suzuki1985decomposition}, we obtain a trapezoidal-state approximation given by $\hat U\approx\hat U_{\beta, K}$~\cite{gunlycke2025guided} such that
	\begin{equation}\label{d_AdiabaticU}
		\hat{U}_{\beta,K} = \exp\!\bigg(\!\!-\!\frac{i\beta\hat H}{2}\bigg)\prod_{k=1}^{K-1}\exp\left[-i\beta\hat{H}(1-\eta_k)\right]\hat{U}_0,
	\end{equation}
	where the operator $\exp(-i\beta\hat{H}_0/2)$ has been dropped because it affects only the global phase when acting on $|\Phi_0\rangle$. The trapezoidal state is expected to be close to the ground state under the following inequalities:
	\begin{equation}\label{Conditions}
		\beta \ll(\hbar\omega_{\text H})^{-1}\le(\hbar\omega_{\text L})^{-1}\ll\beta K,
	\end{equation}
	where $\hbar\omega_\text{L}$ is the energy gap between the first excited state and the ground state of $\hat{H}_0$, and $\hbar\omega_\text{H}$ is the energy gap between the highest excited state and the ground state of $\hat{H}_0$. Equation~\eqref{Conditions} provides a heuristic guideline for the trapezoidal state to reach adiabaticity. For more details, see Appendix~\ref{s.A_1}.
	
	To represent the trapezoidal state on a quantum register, we further need to expand the Hamiltonian $\hat H(\eta)=\sum_{r}c_r(\eta)\bigotimes_{q\in\mathcal Q}\hat\sigma_{r_q}$ in terms of tensor products of Pauli operators $\hat\sigma_{r_q}$, where $c_r(\eta)$ are Pauli coefficients and $\mathcal Q$ is an index set of qubits in the register.  We can then apply the Lie-Trotter approximation~\cite{suzuki1985decomposition,huyghebaert1990product}, which yields
	\begin{equation}\label{LieTrotter}
		\exp\left[-i\beta\hat H(\eta)\right]\approx\prod_r \exp\left[-i\beta c_r(\eta)\bigotimes_{q\in\mathcal Q}\hat\sigma_{r_q}\right].
	\end{equation} 
	Each factor on the right-hand side is then decomposed into a component circuit. Together, these components for all indices $k$ and $r$ in Eqs.~\eqref{d_AdiabaticU} and \eqref{LieTrotter} form a complete circuit that produces our guiding state $|\Psi_0\rangle$.
	
	\subsection{Electronic-structure calculations}\label{s.2_3}
	The atomic coordinates along the minimum-energy path for reaction~\eqref{reaction} were computed using the nudged elastic band (NEB) method~\cite{mills1995reversible,jonsson1998nudged}. Nonrelativistic, finite-basis electronic energies for structures along an NEB-computed minimum-energy path with 63 replicas were computed using the unrestricted coupled-cluster singles and doubles (CCSD) method with the valence quadruple-zeta polarization (def2-QZVP)\cite{weigend2005balanced} Gaussian-type basis set. In addition, unrestricted Hartree-Fock energies were computed with the def2-QZVP and Slater-type orbitals with 6 primitive Gaussian-type orbitals (STO-6G)~\cite{hehre1969self} basis sets to provide basis-set corrections as described below.  All classical electronic-structure calculations were performed using \texttt{PSI4}~\cite{smith2020psi4}, and NEB simulations were conducted using the \texttt{ASE}~\cite{larsen2017atomic} interface to \texttt{PSI4}.
	
	\subsection{Basis-set correction}\label{s.2_4}	
	In this work, we use the minimal basis, STO-6G~\cite{hehre1969self}, in CVQE calculations to preserve the one-to-one correspondence between HF spin orbitals and qubits. This allows us to directly use the Jordan-Wigner transformation~\cite{jordan1928paulische} and minimize the number of required qubits. To directly compare the energies computed with the minimal STO-6G basis and the larger def2-QZVP basis, we introduce the basis-set correction energy
	\begin{equation}\label{basis_set_correction}
		E_{\text{bsc}} = E_{\text{HF/def2-QZVP}}-E_{\text{HF/STO-6G}}
	\end{equation}
	and add it to the calculated energies using the STO-6G basis.
	
	\section{Results and Discussion}\label{s.3}
	\subsection{Molecular structures along the reaction path}\label{s.3_1}
	\begin{figure}
		\includegraphics[scale=0.6]{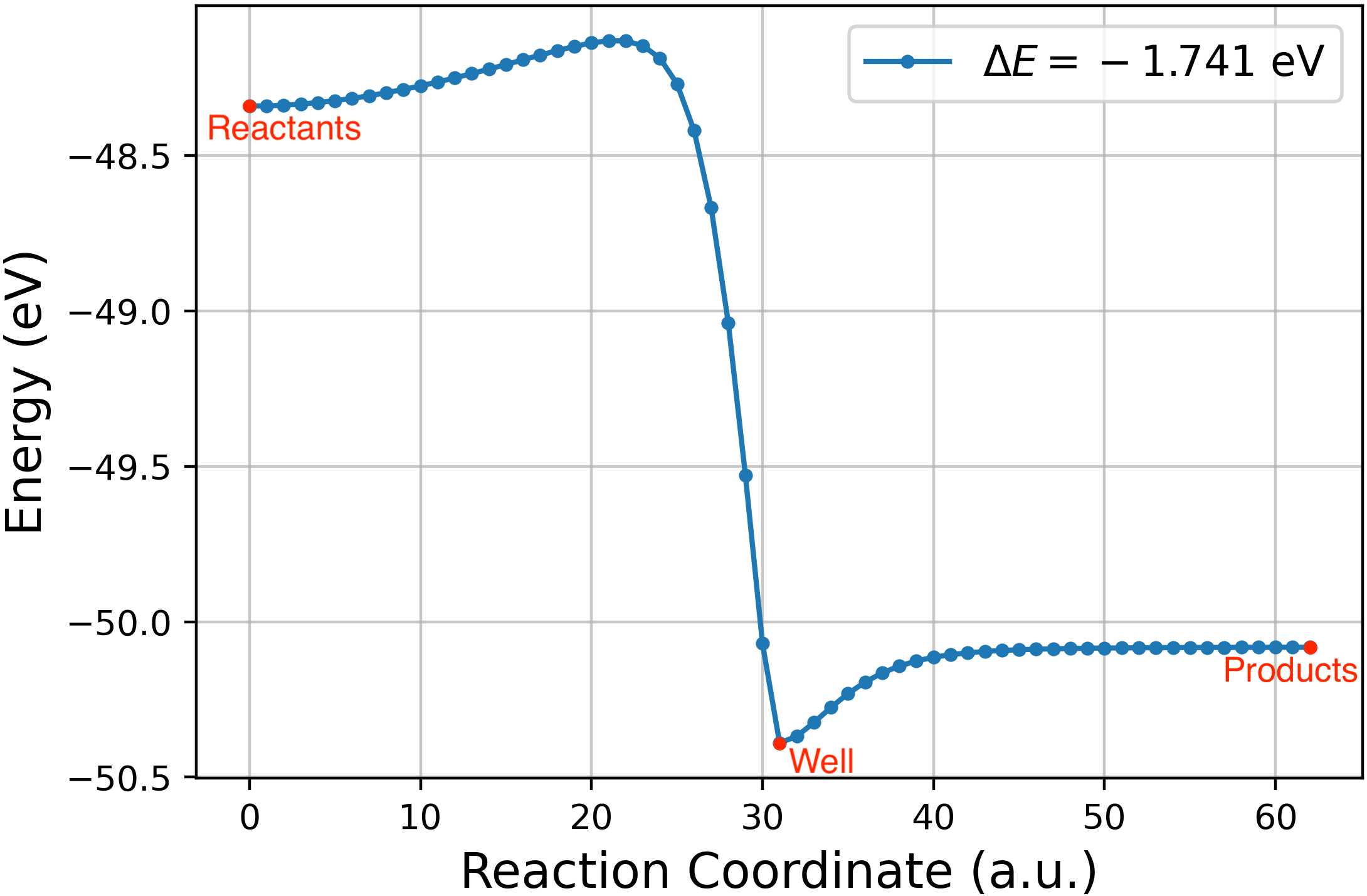}
		\caption{Electronic energies along the minimum-energy path for reaction~\eqref{reaction} computed by the NEB method using CCSD/def2-QZVP. Energies corresponding to the reactant, collision, and product complexes are highlighted in red at coordinates 0, 31, and 62, respectively.}
		\label{fig:NEB_energy}
	\end{figure}
	
	\begin{figure*}
		\includegraphics[scale=0.55]{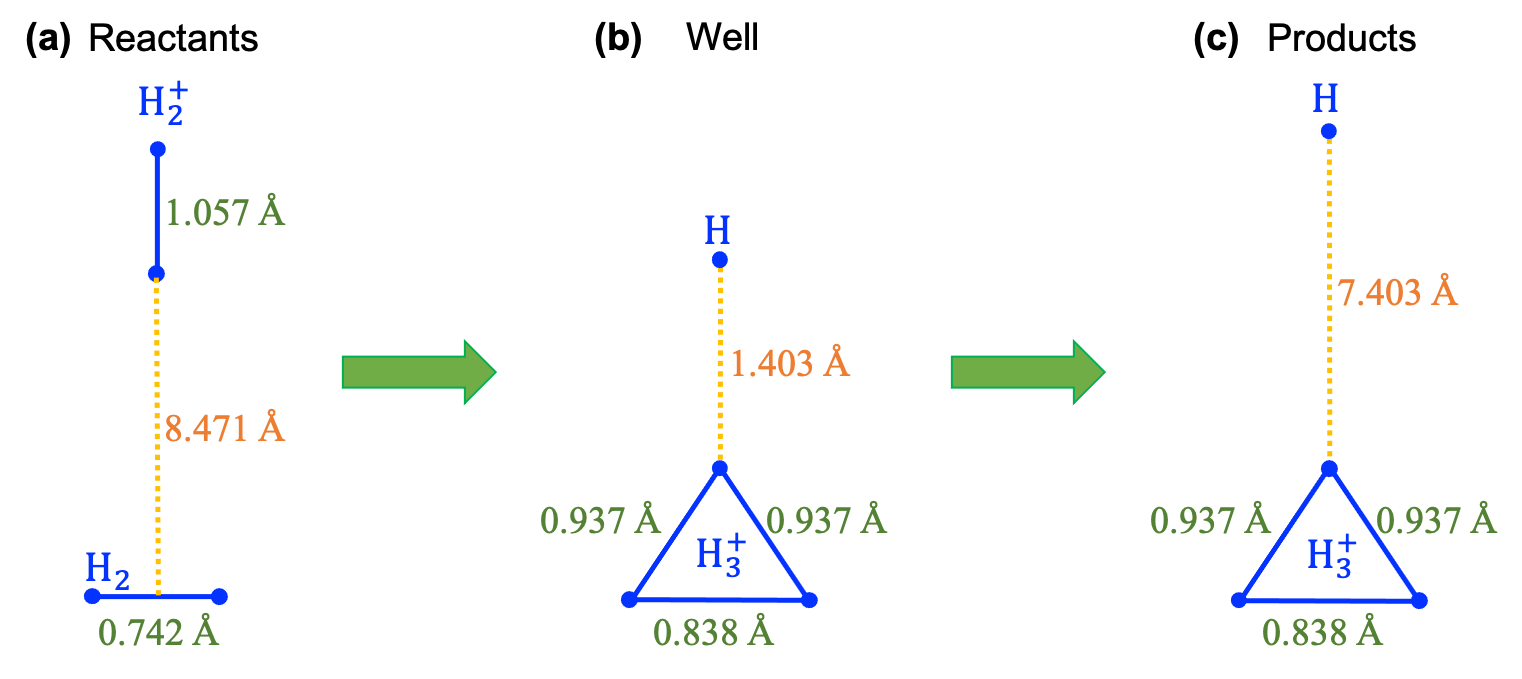}
		\caption{Representative molecular structures along the minimum-energy path for reaction~\eqref{reaction}. Hydrogen atoms are shown in blue and interatomic distances are shown in green and orange. Shown are the geometries of the (a) reactants ($\ce{H_2^+}$ and $\ce{H_2}$ are separated), (b) well (the system is at the lowest energy during the reaction), and (c) products ($\ce{H_3^+}$ and $\ce{H}$ are separated).}
		\label{fig:MoleGeo}
	\end{figure*}
	
	The results of the NEB calculations are summarized in Figs.~\ref{fig:NEB_energy} and \ref{fig:MoleGeo}. Consistent with prior studies on this reaction~\cite{krenos1976crossed,stine1978charge,pollard1991state}, our simulations at the CCSD/def2-QZVP level predict a small barrier along the minimum-energy path followed by a steep decrease in energy leading to a stable collision complex. The overall reaction is found to be exothermic, with an energy release of $-1.741$\,eV, in excellent agreement with the experimental value of $-1.74$\,eV~\cite{oka2013interstellar} (see Table~\ref{tab: reaction_E}).  The molecular structures of the reactant complex, the collision complex (hereafter referred to as the ``well''), and the product complex are shown in Fig.~\ref{fig:MoleGeo}. The full list of atomic coordinates is given in Appendix~\ref{s.A_2}. Geometries from our NEB calculations are then used for the subsequent CVQE calculations.
	\begin{table}[ht]
		\centering
		\caption{Reaction energies from quantum (CVQE) and classical (HF, CCSD, FCI) methods compared to the experimental value in Ref.~\cite{oka2013interstellar}.}
		\label{tab: reaction_E}
		\begin{threeparttable}
			\begin{tabular}{|c|c|c|}
			\hline
			$\qquad\quad$Method$\qquad\quad$ & $\qquad\quad$Basis$\qquad\quad$ & $\qquad\Delta E$ (eV)$\qquad$\\
			\hline
			HF & def2-QZVP & $-1.659$ \\
			\hline
			CCSD & def2-QZVP & $-1.741$\\
			\hline
			FCI & STO-6G\tnotex{a}  & $-1.804$ \\
			\hline
			CVQE & STO-6G\tnotex{a}  & $-1.804$ \\
			\hline
			Experiment & - & $-1.74$ \\
			\hline
			\end{tabular}
			\begin{tablenotes}
				\item[a] \label{a} STO-6G refers to STO-6G plus the correction of Eq.~\eqref{basis_set_correction}.
			\end{tablenotes}
		\end{threeparttable}
	\end{table}
	
	\subsection{CVQE energies along the reaction path}\label{s.3_2}
\begin{figure}
		\includegraphics[scale=0.6]{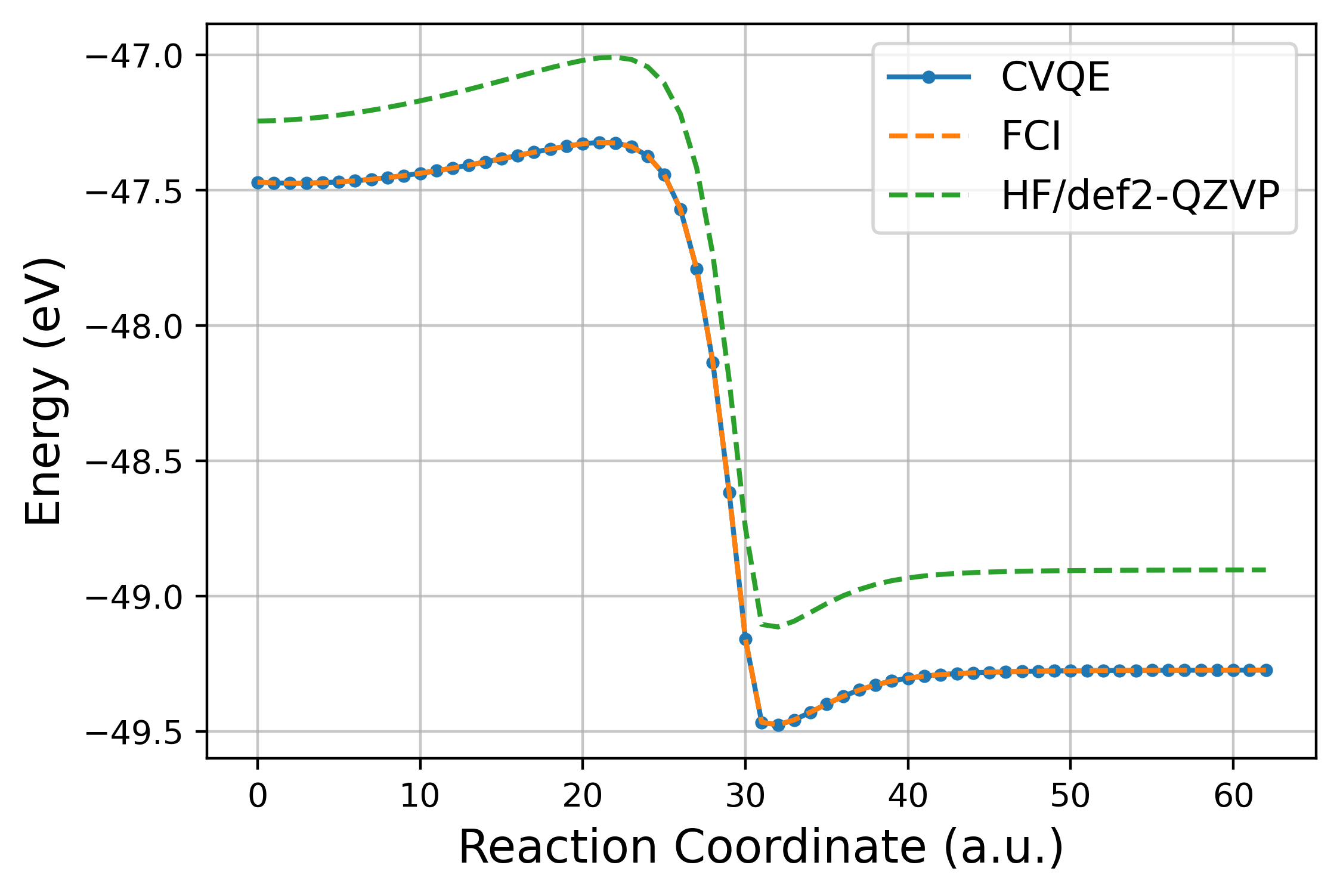}
		\caption{Electronic energies along the minimum-energy path for reaction \eqref{reaction} with the reactant, well, and product states at coordinate 0, 31, and 62, respectively. The reactant-product energy differences are $\Delta E=-1.804$\,eV for CVQE and FCI, and $\Delta E=-1.659$\,eV for HF/def2-QZVP.}
		\label{fig:E_vs_IRC}
	\end{figure}

	Figure~\ref{fig:E_vs_IRC} presents the ground-state energies obtained with the CVQE method using \texttt{Qiskit}'s~\cite{qiskit2024} noiseless simulator as a blue solid curve, which is close to the orange dashed curve obtained with classical full configuration interaction (FCI). Both sets of \legendfont{CVQE} and \legendfont{FCI} calculations use the STO-6G basis set and the correction $E_{\text{bsc}}$ in Eq.~\eqref{basis_set_correction}. For comparison, HF energies with the larger basis set def2-QZVP are shown by the green dashed curve. Both the \legendfont{CVQE} and \legendfont{FCI} calculations yield the same reaction energy, $\Delta E=-1.804$\,eV.
	
	\subsection{Probability distributions of each CVQE stage}\label{s.3_3}
	To understand the impact of different CVQE stages on the state, we analyze the probability distributions at each stage and compare them with the ground-state probability distribution, \legendfont{FCI-pGndD}. Specifically, we investigate: the trapezoidal-state distribution \legendfont{CVQE-pTD}, the guiding-state distribution \legendfont{CVQE-pGD}, the distribution of guiding-state measurement outcomes \legendfont{CVQE-sGD}, and the optimized-state distribution \legendfont{CVQE-pOD}. The FCI ground state of the system is a superposition of 12 out of the 256 Fock states and has a total energy $E_\text{g}\approx -47.555$\,eV for the STO-6G basis set at the well geometry. Although these 12 states are characterized by a point group analysis in Appendix~\ref{s.A_3}, the CVQE calculations have not been constrained and cover the full 256-dimensional Fock space. Comparing the various probability distributions for the most important Fock states contributing to the ground state is therefore a good approach to gain insight into the CVQE process at each stage and to select the guiding-state parameters $\beta$ and $K$.
	
	We have identified three different parameter regimes as they relate to the inequalities $\beta \ll(\hbar\omega_{\text H})^{-1}\le(\hbar\omega_{\text L})^{-1}\ll\beta K$ in Eq.~\eqref{Conditions}, with $(\hbar\omega_\text{H})^{-1}\approx 0.806$\,$(\text{hartree})^{-1}$ and $(\hbar\omega_\text{L})^{-1}\approx 2.387$\,$(\text{hartree})^{-1}$ being the computed model (i.e., Hartree-Fock) gaps in the well geometry.
	
	\subsubsection{Adiabatic guiding state}\label{s.3_3_1}
	The parameters $\beta=0.1$\,$(\text{hartree})^{-1}$ and $K=500$ were chosen such that the inequalities in Eq.~\eqref{Conditions} are satisfied at both ends by a healthy margin, while maximizing the number of steps. This combination yielded an energy error $E_{\text{error}}$ of $0.001$\,eV for the produced trapezoidal state and $0.009$\,eV for the guiding state relative to $E_\text{g}$, both well below the chemical accuracy energy $0.043$\,eV.
	
\begin{figure}[t]
		\includegraphics[scale=0.6]{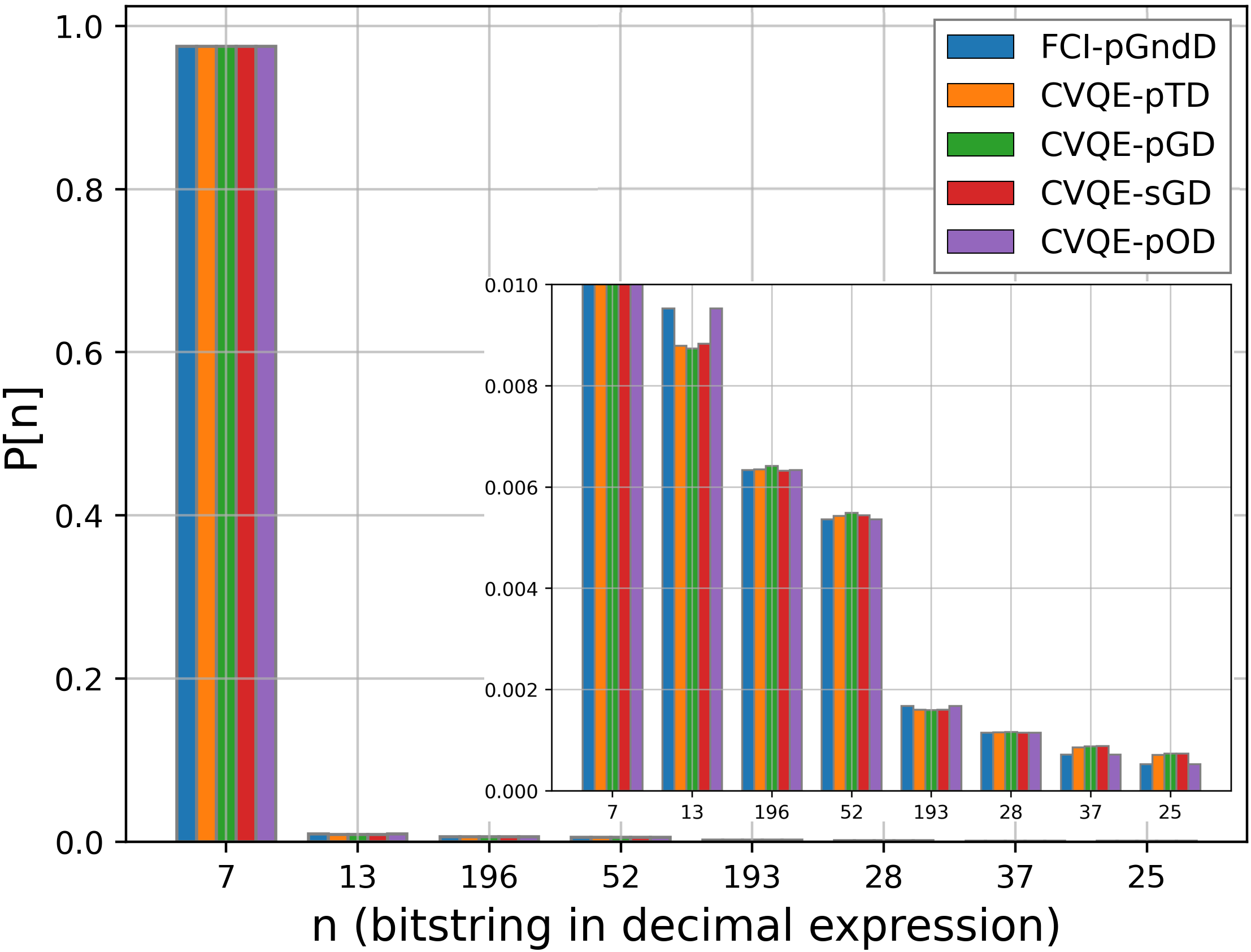}
		\caption{Probability distributions for the well geometry over the eight dominant symmetry-allowed Fock states $|n\rangle$ in the decreasing order obtained on a noiseless simulator. The inset shows an enlarged scale of the main panel. The trapezoidal state was prepared with $K=500$ steps and $\beta=0.1$\,$(\text{hartree})^{-1}$, and the guiding state was sampled using $|\mathcal S|=10^6$ shots. With the inequalities in Eq.~\eqref{Conditions} satisfied, small Lie-Trotter error, and sufficient sampling, the distributions are in relatively good alignment. The energy error of the optimized state is negligible.}
		\label{fig:P_combo_2b}
	\end{figure}

\begin{figure}[t]
		\includegraphics[scale=0.6]{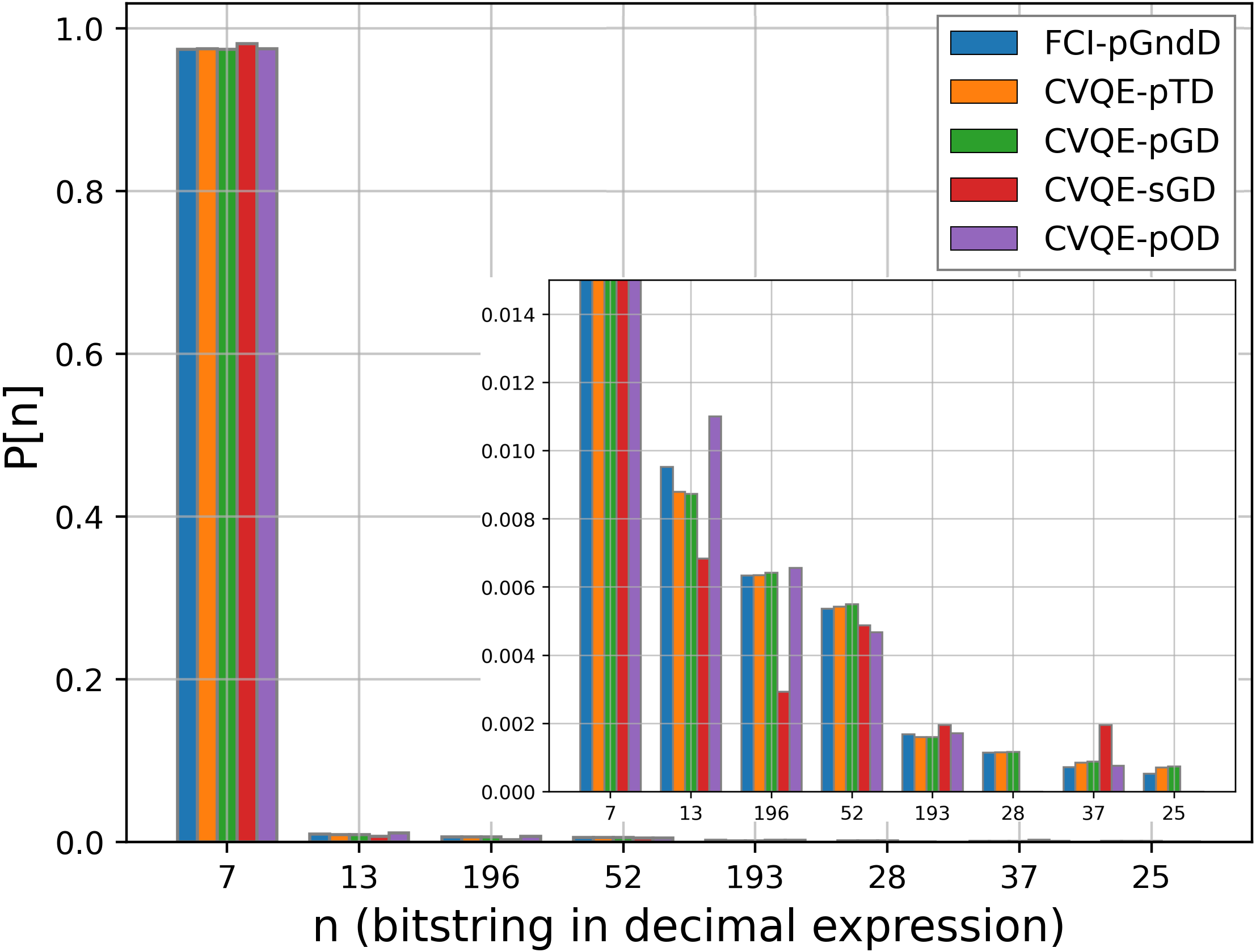}
		\caption{Probability distributions for the well geometry over the eight dominant symmetry-allowed Fock states $|n\rangle$ in the decreasing order obtained on a noiseless simulator. The inset shows an enlarged scale of the main panel. The trapezoidal state was prepared with $K=500$ steps and $\beta=0.1$\,$(\text{hartree})^{-1}$, and the guiding state was sampled using $|\mathcal S|\approx10^3$ shots. The inequalities in Eq.~\eqref{Conditions} are satisfied and the Lie-Trotter error is small, but the small sample size causes the sampled distribution \legendfont{CVQE-sGD} to deviate from the guiding-state distribution \legendfont{CVQE-pGD}. The distribution of the optimized state \legendfont{CVQE-pOD} recovers sufficiently to limit the energy error to $E_{\text{error}}\approx 0.068$\,eV, which is close to chemical accuracy.}
		\label{fig:P_combo_5b}
	\end{figure}

	As one might expect given the low energy error, the associated probability distributions \legendfont{CVQE-pTD} and \legendfont{CVQE-pGD}, shown as orange and green bars respectively in Figs.~\ref{fig:P_combo_2b} and ~\ref{fig:P_combo_5b}, align well with the blue bars for \legendfont{FCI-pGndD}.	
	
	Figure~\ref{fig:P_combo_2b} also shows that the number of shots, $|\mathcal S|=10^6$, is large enough for the sampled distribution \legendfont{CVQE-sGD} and consequently the optimized distribution \legendfont{CVQE-pOD} to match the other distributions. In this case, the energy error of the optimized state is only $6\times 10^{-14}$\,eV.  In contrast, in Fig.~\ref{fig:P_combo_5b}, if the number of shots is restricted to $|\mathcal S|=2^{10}$ (approximately $10^3$), the sampling distribution \legendfont{CVQE-sGD} deviates significantly from the guiding state \legendfont{CVQE-pGD}. This also causes a suboptimal optimized distribution \legendfont{CVQE-pOD} and an energy error of $0.068$\,eV just outside of chemical accuracy.
	
	To understand the sampling breakdown, consider that the probability of finding specific Fock states within samples depends primarily on the number of shots $|\mathcal S|$. A state $n$ of probability $p_n$ is observed at least once with probability $1-(1-p_n)^{|\mathcal S|}$, with expected count $|\mathcal S|p_n$. Consequently, if $|\mathcal S|<(p_n)^{-1}$, state $n$ is unlikely to be sampled even once. Since the CVQE optimization subspace within the GSA is spanned by the sampled Fock states, a deterioration of the sampling distribution is expected to lead to a drop in the accuracy of the optimized energies $E_{\text{pOD}}$ as observed.
	
	\subsubsection{Ideal trapezoidal state}\label{s.3_3_2}
	\begin{figure}
		\includegraphics[scale=0.6]{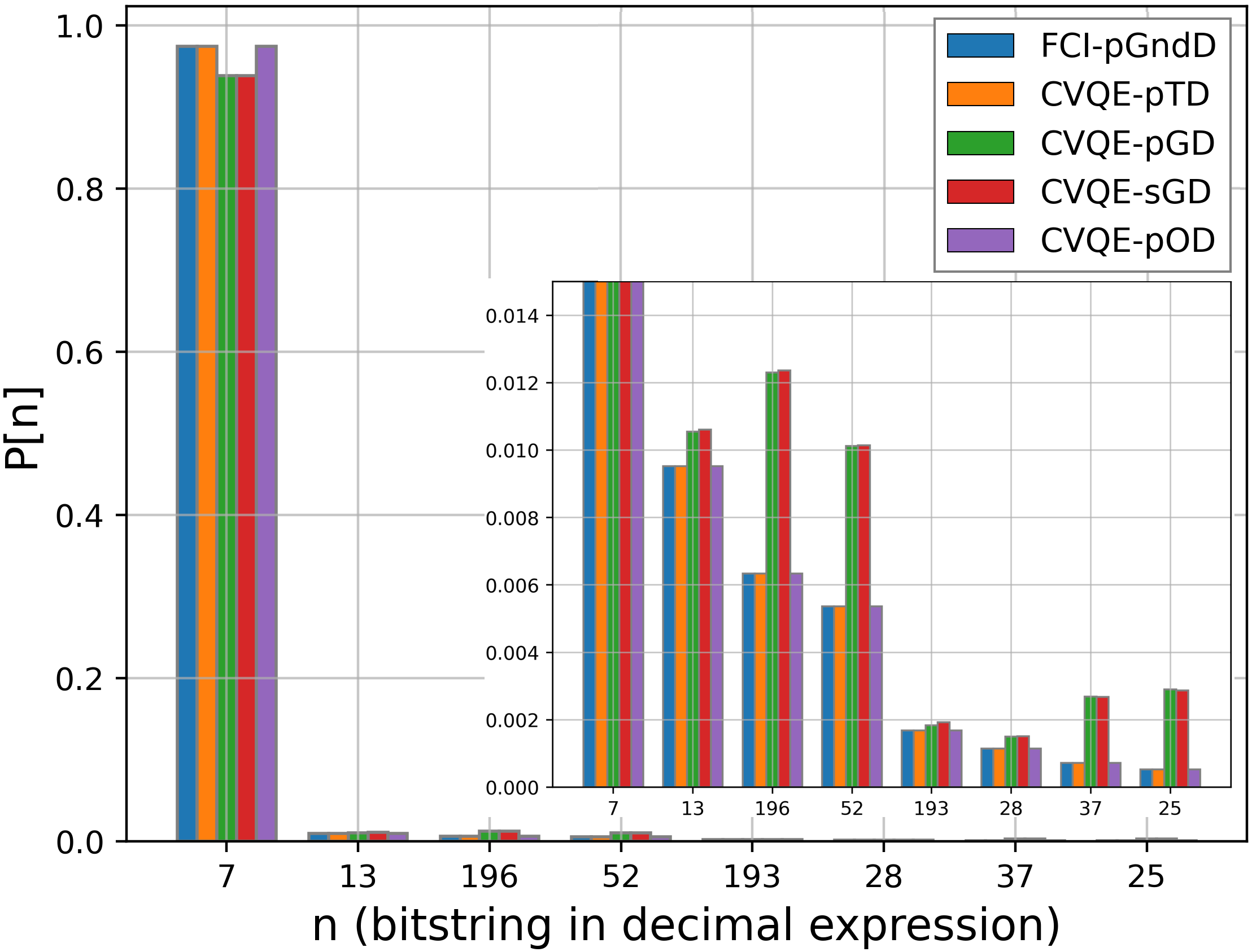}
		\caption{Probability distributions for the well geometry over the eight dominant symmetry-allowed Fock states $|n\rangle$ in the decreasing order obtained on a noiseless simulator. The inset shows an enlarged scale of the main panel. The trapezoidal state was prepared with $K=1000$ steps and $\beta=1.0$\,$(\text{hartree})^{-1}$, and the guiding state was sampled using $|\mathcal S|=10^6$ shots. The large $\beta K$ yields outstanding agreement between the FCI ground-state distribution \legendfont{FCI-pGndD} and the produced trapezoidal-state distribution \legendfont{CVQE-pTD}. The guiding-state distribution \legendfont{CVQE-pGD} and, concomitantly, the sampled distribution \legendfont{CVQE-sGD}, however, are off due to significant Lie-Trotter errors. Given the large number of shots, the final optimized solution is recovered with negligible energy error.}
		\label{fig:P_combo_4b}
	\end{figure}
	
	The two inequalities in Eq.~\eqref{Conditions} are not necessarily of equal importance. The purpose of the left inequality is to prevent excitations during the adiabatic process; a sufficient choice in our calculations for the trapezoidal-state preparation is $\beta=1.0$\,$(\text{hartree})^{-1}$. The advantage of relaxing this inequality slightly is that a larger evolutionary step size $\beta$ facilitates a larger $\beta K$ for the same circuit depth as measured by $K$, improving the right inequality. The distributions for $K=1000$ are shown in Fig.~\ref{fig:P_combo_4b}. We find that the trapezoidal distribution \legendfont{CVQE-pTD} closely matches \legendfont{FCI-pGndD} with the associated trapezoidal energy error $E_{\text{error}}= 4.3\times 10^{-6}$\,eV, which is four orders of magnitude smaller than the chemical accuracy.
	
	Looking at the guiding-state distribution \legendfont{CVQE-pGD} and the associated sampled distribution \legendfont{CVQE-sGD}, we find that these distributions differ significantly from \legendfont{FCI-pGndD}. This indicates that the Lie-Trotter approximation~\cite{suzuki1985decomposition,huyghebaert1990product} in Eq.~\eqref{LieTrotter} has started to fail in this regime. There are two reasons for this: (1) the product-formula error per step is predicted to increase as $\beta^2$ as $\beta$ increases~\cite{childs2021theory, layden2022first}, and (2) this error is compounded by the larger number of steps $K$. In principle, one could implement a higher-order Trotter formula~\cite{suzuki1985decomposition, suzuki1991general} to reduce the Trotter error, but the decomposition would dramatically increase the quantum circuit depth, making this approach impractical until gate errors are significantly reduced.
	
	A certain level of error in the guiding state can be tolerated, however, as the error in the optimized state starts to accumulate once the sampled Fock state set $\mathcal N_{\mathcal S}$ becomes sufficiently distorted. In Fig.~\ref{fig:P_combo_4b}, we see that for $|\mathcal S|=10^6$ shots, the distribution \legendfont{CVQE-pOD} closely matches \legendfont{FCI-pGndD}, demonstrating that the guiding-state error has been largely compensated for by the variational parameter collection $\theta$ during the subsequent classical optimization. In a sense, the CVQE method exhibits, by design, an intrinsic error-resilient feature, which is valuable for NISQ-era computation.
	
	\subsubsection{Single-step guiding state}\label{s.3_3_3}
	\begin{figure}
		\includegraphics[scale=0.6]{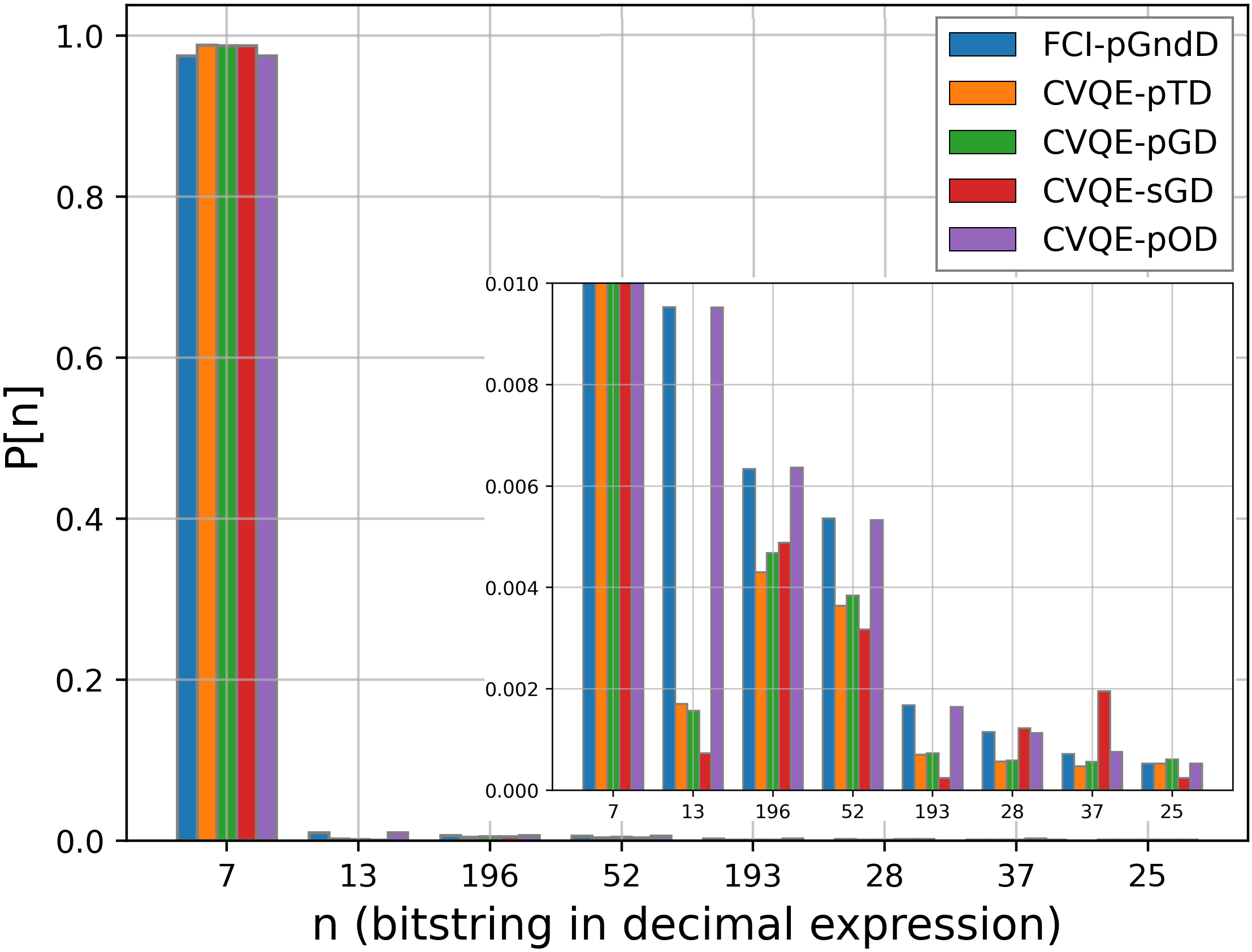}
		\caption{Probability distributions for the well geometry over the eight dominant symmetry-allowed Fock states $|n\rangle$ in the decreasing order obtained on a noiseless simulator. The inset shows an enlarged scale of the main panel. The trapezoidal state was prepared with a single $K=1$ step and $\beta=1.0$\,$(\text{hartree})^{-1}$, and the guiding state was sampled using $|\mathcal S|\approx4\times10^3$ shots. Despite the inequalities in Eq.~\eqref{Conditions} not being met, the sampling of the guiding state identifies the majority of the most important Fock states. Concomitantly, the optimized-state distribution \legendfont{CVQE-pOD} agrees well with the FCI ground-state distribution \legendfont{FCI-pGndD}. The energy error $E_{\text{error}}\approx 8.667\times 10^{-4}$\,eV is also well below the threshold for chemical accuracy.}
		\label{fig:P_combo_8}
	\end{figure}
	
	A challenge as we perform calculations on larger systems is that the number of interactions also increases, generally with the fourth power of the system size. This means that the practical bottleneck in our approach is the circuit depth. One question is then how much we can control the circuit depth using the number of steps $K$. In a separate study, it was found that the intrinsic error-resilient effect in CVQE is sufficient to obtain total energies for hydronium (H$_3$O$^+$) at chemical accuracy, even for a single step, $K=1$~\cite{gunlycke2025guided}. In this limit, neither of the inequalities in Eq.~\eqref{Conditions} can be simultaneously satisfied. We therefore expect the trapezoidal state to be at best a rough approximation of the ground state.
	
	In Fig.~\ref{fig:P_combo_8}, we indeed find that for $\beta=1.0$\,$(\text{hartree})^{-1}$ and $|\mathcal S|=2^{12}$ shots, neither \legendfont{CVQE-pTD} nor \legendfont{CVQE-pGD} accurately reproduces \legendfont{FCI-pGndD}. The sampled distribution, \legendfont{CVQE-sGD}, is, consequently, also not a good representation of the ground-state distribution. Yet we find that the optimized distribution, \legendfont{CVQE-pOD}, largely recovers \legendfont{FCI-pGndD} for all the dominant Fock states. This agreement is reflected in the energy as well, with the optimized energy error $8.667\times 10^{-4}$\,eV almost two orders of magnitude below chemical accuracy.
	
	In the single-step regime, the parameter $\beta$ through the unitary operator $\exp(-i\beta\hat{H}/2)$ from Eq.~\eqref{d_AdiabaticU} becomes a control knob for the evolution away from the Hartree-Fock ground state $|\Phi_0\rangle$. Although this effect was observed through the total energy~\cite{gunlycke2025guided}, we can also see it in the probability distributions directly. In Fig.~\ref{fig:P_QC_Ansatz_dt}, we show the guiding-state distribution, \legendfont{CVQE-pGD}, as a function of this parameter. To prepare for QPU calculations with a reduced circuit depth, we drop all system interactions less than 0.02 hartree. Additionally, we drop diagonal Pauli terms, which do not affect the guiding state if executed first. Note that in the single-step regime, the Lie-Trotter error does not compound, so the guiding state, which is ultimately the measured state, is a reasonable representation of the trapezoidal state. As expected, we observe a monotonic decrease of the Hartree-Fock state component ($n=7$) and a monotonic increase of the other significant Fock states as $\beta$ increases.
	
	\subsubsection{Quantum Processing}\label{s.3_3_4}
	\begin{figure}
		\includegraphics[scale=0.61]{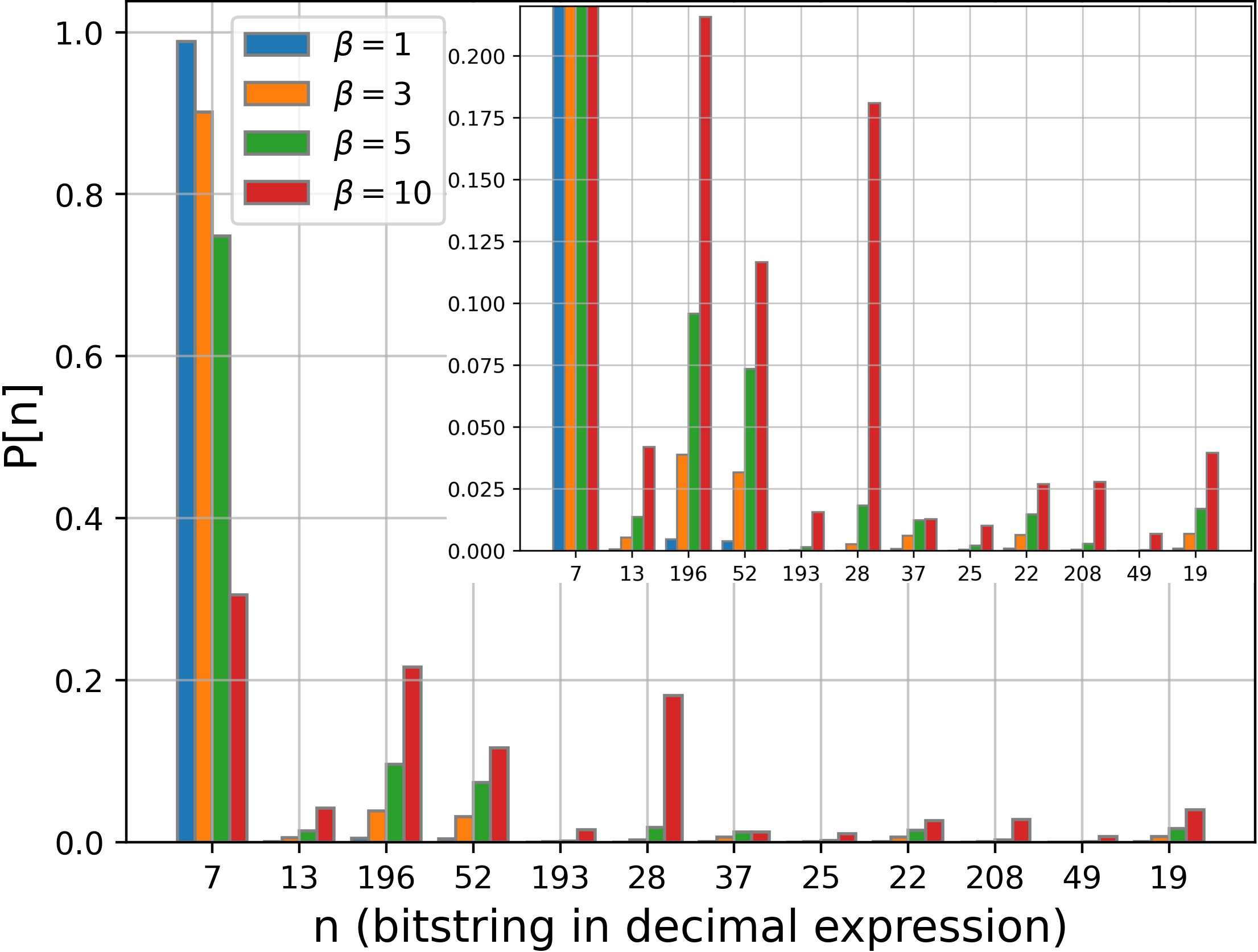}
		\caption{The guiding-state distribution \legendfont{CVQE-pGD} of the well geometry over the Fock states $|n\rangle$. In addition to using a single $K=1$ step of variable $\beta$ [in units of $(\text{hartree})^{-1}$], Hamiltonian interactions with a magnitude below $0.02$ hartree are dropped, along with diagonal Pauli terms, to limit the circuit depth. The inset shows an enlarged scale of the main panel.}
		\label{fig:P_QC_Ansatz_dt}
	\end{figure}
	
	\begin{figure*}
		\includegraphics[scale=0.48]{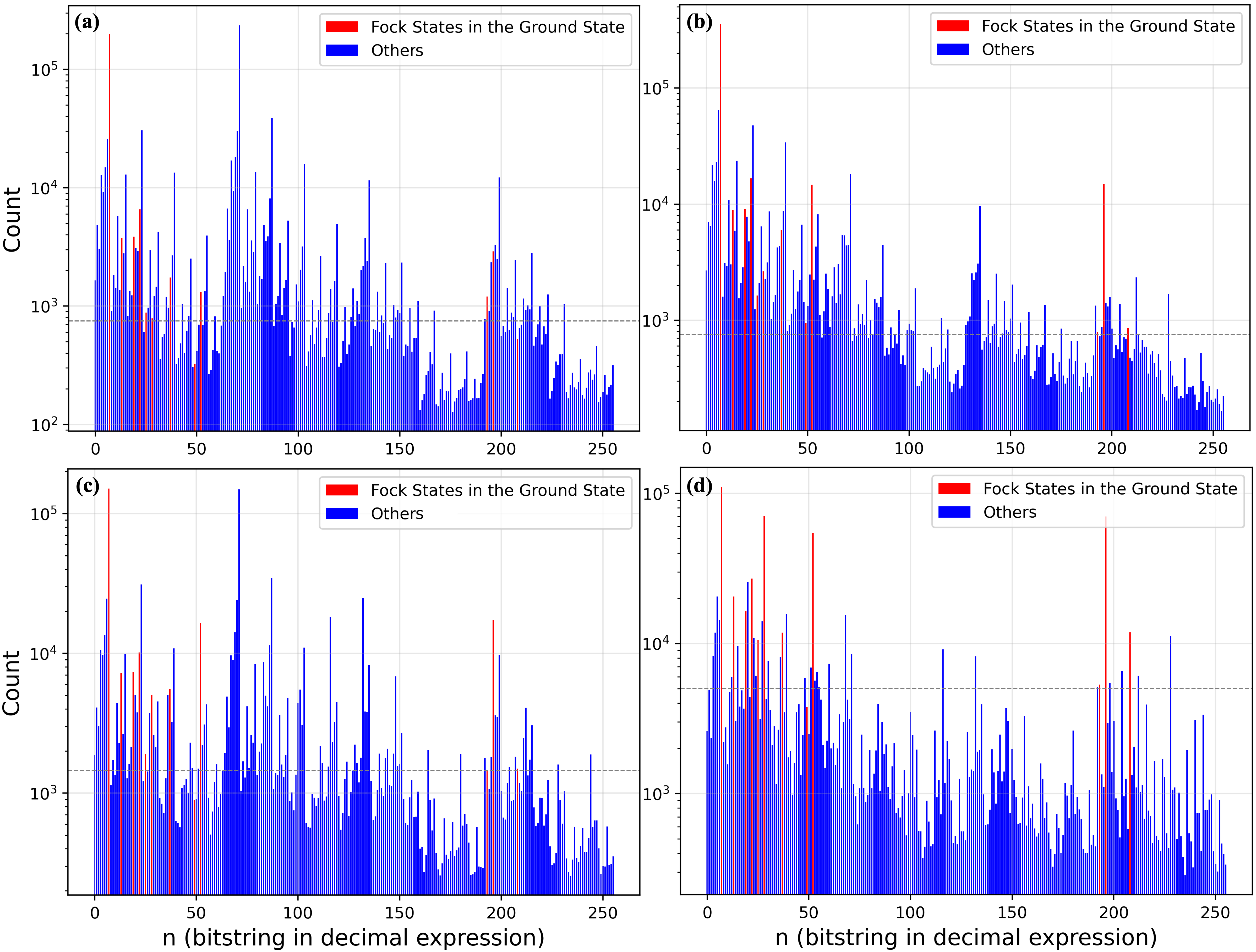}
		\caption{Sampling counts from the QPU, \textit{ibm\_aachen}, across all 256 Fock states $|n\rangle$ for the well geometry, with $|\mathcal S|=10^6$. The red bars represent the Fock states that appear in the ground state, while the blue bars represent those that do not. The gray dashed lines in each panel are the count thresholds. \textbf{(a)} $\beta=1$\,$(\text{hartree})^{-1}$: Of the 128 Fock states above a count threshold of 750, 31 reside within the three-electron subspace. SNR $\approx 0.078$. \textbf{(b)} $\beta=3$\,$(\text{hartree})^{-1}$: Of the 129 Fock states above a count threshold of 750, 38 reside within the three-electron subspace. SNR $\approx 0.093$. \textbf{(c)} $\beta=5$\,$(\text{hartree})^{-1}$: Of the 110 Fock states above a count threshold of 1450, 31 reside within the three-electron subspace. SNR $=0.1$. \textbf{(d)} $\beta=10$\,$(\text{hartree})^{-1}$: Of the 39 Fock states above a count threshold of 5000, 11 reside within the three-electron subspace. SNR $=0.282$.}
		\label{fig:Counts_QPU}
	\end{figure*}
	
	Up to this point, the presented results were obtained using a noiseless IBM simulator. To see how real-world system noise impacts our calculations, we also performed calculations on the \textit{ibm\_aachen} quantum computer. The circuit depth was again kept minimal by choosing a single-step guiding state, $K=1$, dropping system interactions less than 0.02 hartree, and dropping diagonal Pauli terms, as discussed. This process reduced the number of CNOT gates in the final circuit to 152. To obtain an accurate sampling distribution, we used $10^6$ shots. The number of shots is much larger than the total number of Fock states, which is impractical and nonscalable for larger systems. To better understand how CVQE behaves for large systems, we built our subspace out of a reduced sample set that includes only Fock states with counts above a count threshold, indicated by the gray dashed line in Fig.~\ref{fig:Counts_QPU}. This threshold is tuned high enough to capture dominant symmetry-allowed states while suppressing symmetry-forbidden states as an error-mitigation strategy. The resulting subspace is further restricted to three electrons for diagonalization, ensuring that all cases in Fig.~\ref{fig:Counts_QPU} achieve chemical accuracy. Energy convergence remains independent of the threshold choice, provided all symmetry-allowed states are retained and the resulting subspace is computationally manageable for diagonalization.
	
	The sampled probability distributions, \legendfont{CVQE-sGD}, for different $\beta$ values are shown in Fig.~\ref{fig:Counts_QPU}. Note that due to the noise in these calculations, the guiding-state probability distributions, \legendfont{CVQE-pGD}, can be spread out over all 256 Fock states. This spreading even occurs for $\beta=1$\,$(\text{hartree})^{-1}$, for which the probability distribution is supposed to be $98.9\%$ in the Hartree-Fock state based on the noiseless calculation shown in Fig.~\ref{fig:P_QC_Ansatz_dt}. We therefore conclude that Fig.~\ref{fig:Counts_QPU}(a) is effectively a noise spectrum, which also explains why the symmetry-allowed probabilities shown in red, with the exception of the leftmost $n=7$ Hartree-Fock state, do not stand out from the symmetry-forbidden states in blue.
	
	As we increase the evolutionary step size to $\beta=10$\,$(\text{hartree})^{-1}$ and the guiding-state probability distribution concomitantly spreads out over the dominant allowed Fock states in Fig.~\ref{fig:P_QC_Ansatz_dt}, we start to see a contrast, with the symmetry-allowed red bars receiving significantly more measured shot counts than the surrounding symmetry-forbidden blue bars. This is especially noticeable in Fig.~\ref{fig:Counts_QPU}(d), for which the expected Hartree-Fock state percentage drops to $30.5\%$ (see Fig.~\ref{fig:P_QC_Ansatz_dt}). This demonstrates that we can use the guiding-state parameter $\beta$ to tune the signal-to-noise ratio (SNR), defined by the ratio of the number of symmetry-allowed states to the number of symmetry-forbidden states, both counted above the threshold in  Fig.~\ref{fig:Counts_QPU}. In fact, the SNR listed in Fig.~\ref{fig:Counts_QPU} increases as $\beta$ increases.
	
	This finding means that the best guiding state in practice is not necessarily the one closest to the ground state, but one that is close enough while also creating enough contrast with the noise background. In this case, we find that for $\beta=10$\,$(\text{hartree})^{-1}$, a SNR of 0.282 and 11 dominant Fock states are correctly identified by our calculation with $|\mathcal S|=10^6$ shots and a count threshold of 5000 in a three-electron subspace. Consequently, the subsequent classical optimization produces a high-quality ground-state approximation with an energy error of $E_{\text{error}}\approx 1.469\times 10^{-4}$\,eV, which is roughly a factor of 300 below chemical accuracy.
	
	\section{Conclusion}\label{s.4}
	We moved beyond traditional energy-convergence benchmarks by introducing a probability distribution analysis capable of pinpointing stage-specific state errors and tracking their propagation throughout the entire optimization process. We analyzed the effects of the trapezoidal-state and guiding-state preparation used in the CVQE by varying the number of steps $K$ and the evolutionary step size $\beta$ across three regimes and characterizing probability distributions at each stage. While the adiabatic guiding states and ideal trapezoidal states approximate the ground-state distribution more closely with $(\hbar\omega_{\text L})^{-1}\ll\beta K$, they also require much deeper quantum circuits, making them more susceptible to quantum noise. Thus, we found that the single-step guiding-state approach works better in practice and allows for calculations surpassing chemical accuracy. This success stems from the intrinsic error resilience of CVQE, evidenced by its capacity to restore optimized state distributions from corrupted earlier-stage probabilities, provided they remain nonzero.
	
	We found that the best calculated predictions for the well-geometry hydrogen bimolecular system on current superconducting-circuit NISQ computers are obtained when we (1) use a single-step guiding state, (2) prune small interactions in the creation of the guiding state to reduce the circuit depth and limit quantum decoherence, (3) use a $\beta$ large enough to create signal-to-noise contrast, and (4) use enough shots to pick up sufficient signal. These findings are consistent with similar calculations on a trapped-ion NISQ computer~\cite{gunlycke2025guided}. Our results provide actionable insights for developing sampling-based quantum algorithms, demonstrating that shallower, less adiabatic circuits yield superior performance by enhancing signal contrast against the hardware noise floor. Ultimately, while these guidelines dictate how we currently balance circuit depth, state quality, sampling overhead, and signal-to-noise ratios, these strategies must remain flexible as quantum computing technology evolves.
	
	Our calculations demonstrate that using the CVQE algorithm with a GSA, we can obtain ground-state energy estimates for the minimum-energy structure within the ion-neutral bimolecular reaction, $\ce{H_2 + H_2^+\rightarrow H_3^+ + H}$, well below chemical accuracy. The success of this approach suggests a transferable methodology that may be applicable to a broad class of similar bimolecular systems. Armed with these design principles, we intend to scale GSA circuits to larger systems tailored for near-term NISQ hardware, accelerating the transition of quantum simulation research toward challenging, large-scale many-electron problems.\\
	
	\textbf{Acknowledgment}: This work was supported by the Office of Naval Research through the U.S. Naval Research Laboratory. We acknowledge QC resources from IBM through a collaboration with the Air Force Research Laboratory (AFRL). Y.-H. L. thanks the National Research Council Research Associateship Programs.
	
	\bibliography{references}
	
	\clearpage
	\begin{center}
		{\bf\large Appendix}
	\end{center}
	\setcounter{section}{0}
	\setcounter{secnumdepth}{3}
	\setcounter{equation}{0}
	\setcounter{figure}{0}
	\renewcommand{\thefigure}{A\arabic{figure}}
	\newcommand\Scite[1]{[S~\citealp{#1}]}
	\makeatletter \renewcommand\@biblabel[1]{[S#1]} \makeatother
	
	\section{Heuristic Derivation of Eq. (10) for Discretized Adiabatic Evolution}\label{s.A_1}
	Here we provide a heuristic derivation for $\Delta t\ll(\omega_\text{H})^{-1}\le(\omega_\text{L})^{-1}\ll T$ (equivalent to $\beta \ll(\hbar\omega_{\text H})^{-1}\le(\hbar\omega_{\text L})^{-1}\ll\beta K$ in Eq.~\eqref{Conditions} of the main text) that is adopted for the discretized adiabatic state preparation. Assuming a nondegenerate spectrum and a weak perturbation that does not close the gap during the evolution, first-order perturbation theory is used to estimate the time scales required for (i) the continuous adiabatic ramp and (ii) its time discretization. The primary references are pages 474-475 and 481 in \cite{shankar1994principles}.
	\subsection{First-order Time-dependent Perturbation Theory}
	We start with 1$^{\text{st}}$-order perturbation theory with the time-dependent Hamiltonian
	\begin{equation}\label{firstH}
		H(t)=H^0+H^1(t)
	\end{equation}
	with $|n^0\rangle$ being the known eigenstates of the non-perturbed Hamiltonian $H^0$, and $H^1(t)$ being the time-dependently perturbed Hamiltonian. Since the basis $\{|n^0\rangle\}$ is complete, we can always express the wave function generally as
	\begin{equation}\label{completeBasis}
		|\psi(t)\rangle = \sum_n c_n(t) |n^0\rangle.
	\end{equation}
	If $H^1(t)$ is absent, then we can find the wave-function coefficients
	\begin{equation}\label{time_c}
		c_n(t)=c_n(0)e^{-iE_n^0 t/\hbar}
	\end{equation}
	by solving the Schr{\"o}dinger equation: $i\hbar|\dot{\psi}\rangle = H^0|\psi\rangle$. So, Eq.~\eqref{completeBasis} can be written as
	\begin{equation}\label{completeSol}
		|\psi(t)\rangle = \sum_n c_n(0)e^{-iE_n^0 t/\hbar}|n^0\rangle.
	\end{equation}
	Now if $H^1(t)$ is present, the initial coefficients $c_n(0)$ in Eq.~\eqref{completeSol} become time-dependent such that the solution is now
	\begin{equation}\label{t_completeSol}
		|\psi(t)\rangle = \sum_n d_n(t)e^{-iE_n^0 t/\hbar}|n^0\rangle.
	\end{equation}
	Taking Eq.~\eqref{t_completeSol} into the full time-dependent Schr{\"o}dinger equation: $i\hbar|\dot{\psi}\rangle = [H^0+H^1(t)]|\psi\rangle$, we can obtain
	\begin{equation}\label{derive_dn}
		\sum_n\left[i\hbar\dot{d}_n-H^1(t)d_n\right]e^{-iE_n^0 t/\hbar}|n^0\rangle=0.
	\end{equation}
	After projecting Eq.~\eqref{derive_dn} with $\langle f^0|e^{iE_f^0 t/\hbar}$, we get the differential equation of $d_n(t)$:
	\begin{equation}\label{Diff_dn}
		i\hbar\dot{d}_f=\sum_n \langle f^0|H^1(t)|n^0\rangle e^{i\omega_{fn}t}d_n,
	\end{equation}
	where 
	\begin{equation}\label{omega_fn}
		\omega_{fn}=\frac{E_f^0-E_n^0}{\hbar}
	\end{equation}
	is the frequency corresponding to the energy difference of state $|f^0\rangle$ and state $|n^0\rangle$.
	
	Now, let's put some initial condition: $|\psi(t_0)\rangle=|i^0\rangle$ and $d_f(t_0)=\delta_{fi}$, which means the wave function starts with one non-perturbed eigenstate $|i^0\rangle$. Then, the solution of the differential equation in Eq.~\eqref{Diff_dn} with the first-order truncation is
	\begin{equation}\label{dn_t}
		d_f(t)=\delta_{fi}-\frac{i}{\hbar}\int_{t_0}^{t}\langle f^0|H^1(t')|i^0\rangle e^{i\omega_{fi}t'}dt'.
	\end{equation}
	The first-order result is self-consistent provided the correction stays small, $|d_f|\ll 1$, which requires $H^1(t)$ to be weak on the scale of the $H^0$ gaps.
	
	\subsection{Adiabatic-runtime requirement: $T\gg (\omega_\text{L})^{-1}$}\label{sec: adiabatic-runtime}
	With the general solution as Eq.~\eqref{dn_t}, we can apply it to a case where the perturbing fixed Hamiltonian $H^1$ is turned on with the time constant $\tau$:
	\begin{equation}\label{H_tau}
		H(t)=H^0+e^{t/\tau}H^1
	\end{equation}
	for $-\infty\le t\le 0$. This means the characteristic timescale for $H^1$ to affect the system is $\tau$. The goal is to figure out how slow $1/\tau$ needs to be such that the system remains in equilibrium. We consider transitions out of the initial state ($f\neq i$), so
	\begin{equation}\label{dn_integrated}
		\begin{aligned}
			d_f(t=0)&=-\frac{i}{\hbar}\int_{-\infty}^{0}\langle f^0|e^{t'/\tau}H^1|i^0\rangle e^{i\omega_{fi}t'}dt'\\
			&=-\frac{i}{\hbar}\frac{\langle f^0|H^1|i^0\rangle}{\frac{1}{\tau}+i\omega_{fi}}.
		\end{aligned}
	\end{equation}
	If $\frac{1}{\tau}\ll \omega_{fi}$, meaning the ramping rate ($1/\tau$) is much slower than the natural frequency of the system ($\omega_{fi}$, for any pair $f\neq i$), then
	\begin{equation}\label{dn_adiabatic}
		d_f(t=0)=\langle f^0|i\rangle = \frac{\langle f^0|H^1|i^0\rangle}{E_i^0-E_f^0},
	\end{equation}
	which recovers the first-order time-independent nondegenerate result. The energies and states become the solutions of $H(t=0)=H^0+H^1$, the Hamiltonian at the end of adiabatic evolution. Strictly, the perturbation needs to be slower than the lowest transition frequency of $H^0$, i.e., $\frac{1}{\tau}\ll \omega_\text{L}$, so the adiabatic-runtime requirement suffices for all $\omega_{fi}$. In other words, the total adiabatic evolution time $T(\simeq\tau)$ that changes the system needs to be much longer than the largest period of the system $(\omega_\text{L})^{-1}$, i.e., $(\omega_\text{L})^{-1}\ll T$. For an adiabatic evolution where we start with the ground state of $H^0$, $\omega_\text{L}=\min_f\{\omega_{f0}\}$. In this case, the lowest frequency corresponds to the transition between the first-excited state and the ground state.
	
	Since we presume $dt'\rightarrow 0$ in the integration of Eq.~\eqref{dn_integrated}, $(\omega_\text{L})^{-1}\ll T$ follows from the continuous-time integral and does not yet account for finite time-step discretization.
	
	\subsection{Time-step resolution requirement: $\Delta t\ll (\omega_\text{H})^{-1}$}\label{sec: time-resolution}
	Numerical implementation requires discretizing the evolution. We discretize the integration in Eq.~\eqref{dn_integrated} by the change of variables: $t'=-k \Delta t$, where $k$ is an integer and $\Delta t$ is the time step size. Then, Eq.~\eqref{dn_integrated} becomes
	\begin{equation}\label{discretized_dn_integrated}
		\begin{aligned}
			d_f(t=0)&=-\frac{i}{\hbar}\sum_{k=\infty}^{0}\langle f^0|e^{-\frac{k\Delta t}{\tau}}H^1|i^0\rangle e^{-i\omega_{fi}k\Delta t}\Delta t\\
			&=-\frac{i}{\hbar}\langle f^0|H^1|i^0\rangle \sum_{k=0}^{\infty}e^{-(\frac{1}{\tau}+i\omega_{fi})k\Delta t}\Delta t\\
			&=-\frac{i}{\hbar}\langle f^0|H^1|i^0\rangle \frac{1}{1-e^{-(\frac{1}{\tau}+i\omega_{fi})\Delta t}}\Delta t.
		\end{aligned}
	\end{equation}
	If $\Delta t$ is too large, then the discrete sum in Eq.~\eqref{discretized_dn_integrated} does not approximate Eq.~\eqref{dn_integrated} to a good numerical accuracy.
	
	If $\left|\left(\frac{1}{\tau}+i\omega_{fi}\right)\Delta t\right|\ll 1$, then
	\begin{equation}\label{disc_dn_approx}
		\begin{aligned}
			d_f(t=0)&\approx -\frac{i}{\hbar}\langle f^0|H^1|i^0\rangle \frac{1}{1-\left[1-\left(\frac{1}{\tau}+i\omega_{fi}\right)\Delta t\right]}\Delta t\\
			&=-\frac{i}{\hbar}\frac{\langle f^0|H^1|i^0\rangle}{\frac{1}{\tau}+i\omega_{fi}}.
		\end{aligned}
	\end{equation}
	
	With the adiabatic-runtime requirement $\frac{1}{\tau}\ll \omega_{fi}$, Eq.~\eqref{disc_dn_approx} reduces to Eq.~\eqref{dn_adiabatic}, which is the same result as when the continuous adiabatic evolution is considered. Therefore, $\left|\left(\frac{1}{\tau}+i\omega_{fi}\right)\Delta t\right|\ll 1$ is the new inequality we have to keep in the discretized description. This new approximation will become
	\begin{equation}
		\begin{aligned}
			\left|\left(\frac{1}{\tau}+i\omega_{fi}\right)\Delta t\right|&=\left|\frac{1}{\tau}+i\omega_{fi}\right|\Delta t\\
			&=\sqrt{(1/\tau)^2+(\omega_{fi})^2}\Delta t\\
			&\approx \omega_{fi}\Delta t \ll 1
		\end{aligned}
	\end{equation}
	when we take $\frac{1}{\tau}\ll \omega_{fi}$. So, we reach the equivalent time-step resolution requirement $\Delta t\ll \frac{1}{\omega_{fi}}$ for the discretized adiabatic evolution. This inequality ensures that the discretized first-order amplitude reproduces the continuum result; it should be read as a resolution requirement on the oscillatory phase, in addition to any step-size constraint imposed by the specific time-propagator used in the numerical implementation. A stringent criterion for $\Delta t$ is that the time-step resolution needs to be much finer than the fastest oscillation, i.e., $\Delta t\ll (\omega_\text{H})^{-1}$, where $\omega_\text{H}\equiv\max_{f,i}\{\omega_{fi}\}$ in the relevant finite basis. For an adiabatic evolution where we start with the ground state of $H^0$, $\omega_\text{H}=\max_f\{\omega_{f0}\}$.
	
	One remark on $\Delta t\ll (\omega_\text{H})^{-1}$: it is a sufficient but not necessary condition. Even when $\omega_{fi}\Delta t\gtrsim 1$, the discrete sum as of Eq.~\eqref{discretized_dn_integrated} may remain reasonably close to the continuous result for a generic non-resonant phase ($\omega_{fi}\Delta t\neq 2\pi m, \forall m \in \mathbb{Z}$) because alternating phases can partially cancel. The genuine failure occurs at resonance: when $\omega_{fi}\Delta t=2\pi m$, then $e^{-i\omega_{fi}\Delta t}\approx 1$ and $[1-e^{-(\frac{1}{\tau}+i\omega_{fi})\Delta t}]\approx \frac{\Delta t}{\tau}$. For large $\tau$, the transition amplitude $d_f\approx-\frac{i}{\hbar}\langle f^0|H^1|i^0\rangle \tau$ can violate the first-order perturbation condition $|d_f|\ll 1$. Thus, the choice of $\omega_{fi}\Delta t\ll 1$ is a conservative robustness condition that is guaranteed for all phases.
	
	\subsection{Summary}
	Combining Secs.~\ref{sec: adiabatic-runtime} and \ref{sec: time-resolution}, the discretized evolution must satisfy
	\begin{equation}\label{adiabatic_cond}
		\Delta t\ll(\omega_\text{H})^{-1}\le(\omega_\text{L})^{-1}\ll T
	\end{equation}
	when it is evolved adiabatically with each discretized $\Delta t$. Note that the evolutionary step size is defined as $\beta\equiv \Delta t/\hbar$. So, $T=\hbar\beta K$ for $K$ equal steps. Therefore, we can equivalently obtain
	\begin{equation}\label{adiabatic_cond2}
		\beta \ll(\hbar\omega_{\text H})^{-1}\le(\hbar\omega_{\text L})^{-1}\ll\beta K.
	\end{equation}
	
	The derivation assumes a weak perturbation with no gap closing along the path, so satisfying Eqs.~\eqref{adiabatic_cond}--\eqref{adiabatic_cond2} would heuristically generate an adiabatic state. However, violating them does not necessarily preclude adiabaticity. While a common adiabatic condition $\frac{\hbar|\langle f(t)|\dot{H}(t)|i(t)\rangle|}{[E_f(t)-E_i(t)]^2}\ll 1$ for all $f\neq i$ requires the off-diagonal change rate of the Hamiltonian to be much less than the square of the instantaneous gap $[E_f(t)-E_i(t)]$ along the adiabatic path, the non-perturbed frequencies $\omega_\text{L}$ and $\omega_\text{H}$ in Eqs.~\eqref{adiabatic_cond}-\eqref{adiabatic_cond2} serve as good references as they are of the leading order in the first-order perturbation theory. A stricter criterion replaces the non-perturbed frequencies by the instantaneous transition frequencies along the evolution path: $\omega_\text{L}\rightarrow\min_t\min_{f\neq i}\{\omega_{fi}(t)\}$ and $\omega_\text{H}\rightarrow\max_t\max_{f\neq i}\{\omega_{fi}(t)\}$, where $\omega_{fi}(t)=\frac{[E_f(t)-E_i(t)]}{\hbar}$.
	
	\section{Geometry Coordinates in Fig. 3}\label{s.A_2}
	In Tables~\ref{tab: Reactants_Geo}-\ref{tab: Products_Geo}, we list the geometry coordinates corresponding to reactants, well geometry, and products shown in Fig.~\ref{fig:MoleGeo}. The numbers here are angstroms. The very small nonzero $x$ and $z$ values are numerical residuals from the geometry optimization.
	
	\begin{table}[htbp]
		\centering
		\caption{Geometry of Reactants, as in Fig.~\ref{fig:MoleGeo}(a)}
		\label{tab: Reactants_Geo}
		\resizebox{\columnwidth}{!}{
			\begin{tabular}{|c|c|c|c|}
				\hline
				& $x$ & $y$ & $z$ \\\hline
				$\ce{H}$ & 0.000000000000 & 8.528398637950 & 0.000000000000 \\\hline
				$\ce{H}$ & 0.000000000000 & 7.471601362050 & 0.000000000000 \\\hline
				$\ce{H}$ & -0.370880024809 & -1.000000000000 & 0.000000000000 \\\hline
				$\ce{H}$ & 0.370880024809 & -1.000000000000 & 0.000000000000 \\\hline
			\end{tabular}
		}
	\end{table}
	
	\begin{table}[htbp]
		\centering
		\caption{Well geometry, as in Fig.~\ref{fig:MoleGeo}(b)}
		\label{tab: LowestEnergy_Geo}
		\resizebox{\columnwidth}{!}{
			\begin{tabular}{|c|c|c|c|}
				\hline
				& $x$ & $y$ & $z$ \\\hline
				$\ce{H}$ & -0.000000000092 & 1.471548475039 & -0.000000000204 \\\hline
				$\ce{H}$ & 0.000000000553 & 0.068342894392 & -0.000000000201 \\\hline
				$\ce{H}$ & -0.418764663051 & -0.769945684835 & -0.000000000513 \\\hline
				$\ce{H}$ & 0.418764662590 & -0.769945684597 & 0.000000000918 \\\hline
			\end{tabular}
		}
	\end{table}
	
	\begin{table}[htbp]
		\centering
		\caption{Geometry of Products, as in Fig.~\ref{fig:MoleGeo}(c)}
		\label{tab: Products_Geo}
		\resizebox{\columnwidth}{!}{
			\begin{tabular}{|c|c|c|c|}
				\hline
				& $x$ & $y$ & $z$ \\\hline
				$\ce{H}$ & -0.000000000092 & 7.471548475039 & -0.000000000204 \\\hline
				$\ce{H}$ & 0.000000000553 & 0.068342894392 & -0.000000000201 \\\hline
				$\ce{H}$ & -0.418764663051 & -0.769945684835 & -0.000000000513 \\\hline
				$\ce{H}$ & 0.418764662590 & -0.769945684597 & 0.000000000918 \\\hline
			\end{tabular}
		}
	\end{table}
	
	\section{Point Group Analysis}\label{s.A_3}
	This section justifies the 12 symmetry-allowed Fock states in the ground-state distributions shown in Figs.~\ref{fig:P_combo_2b}--\ref{fig:Counts_QPU}, as listed in Table~\ref{tab: Fock_state}. Analyzing the system's point group symmetry is necessary, as characterizing each molecular orbital (MO) by its irreducible representation (IR) clarifies these 12 ground-state Fock states.
	
\begin{figure}
		\includegraphics[scale=0.5]{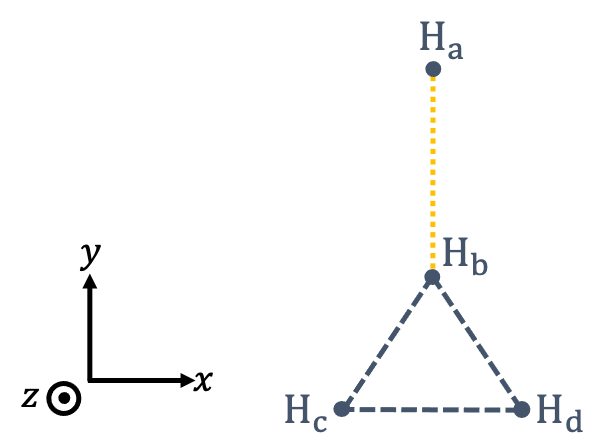}
		\caption{Four-hydrogen geometry on the $x-y$ plane for the convenience of point group analysis. The index of $\ce{H}$ labels each hydrogen atom that corresponds to the description of Table~\ref{tab: transAO}.}
		\label{fig:PG_Geo}
	\end{figure}

	Fixing the four-hydrogen geometries to the $x-y$ plane (i.e., $z=0$) as Fig.~\ref{fig:PG_Geo}, we find the geometry is invariant under the symmetry operators of $C_{2v}$ with its axis along $y$-axis: $E$ (identity), $C_2$ (rotation by $180^\circ$ around the $y$-axis), $\sigma_v(xy)$ (reflection through the $x-y$ plane), and $\sigma_v(yz)$ (reflection through the $y-z$ plane). Therefore, our system belongs to the $C_{2v}$ point group. The transformations of these four atomic orbitals are listed in Table~\ref{tab: transAO}, with each of the hydrogen atoms labeled the same way as in Fig.~\ref{fig:PG_Geo}. Based on the IRs in the character table of $C_{2v}$ [Table~\ref{tab: CharacterTable}], we can construct the Symmetry-Adapted Linear Combinations (SALCs): $\ce{H_a}$, $\ce{H_b}$, and $(\ce{H_c}+\ce{H_d})$ transform as $A_1$, which is symmetric with respect to all operations. $(\ce{H_c-H_d})$ transforms as $B_1$, which is symmetric with respect to $E$ and $\sigma_v(xy)$, but anti-symmetric with respect to $C_2$ and $\sigma_v(yz)$. Because the actual MOs are formed by the linear combinations of SALCs that belong to the same IR, there will be three MOs constructed by the linear combinations of $\ce{H_a}$, $\ce{H_b}$, and $(\ce{H_c}+\ce{H_d})$, and one MO constructed by $(\ce{H_c-H_d})$. So, there will be three MOs of $A_1$ type and one MO of $B_1$ type, i.e., $3A_1+B_1$. The number of a specific IR for the whole system can also be verified by the decomposition formula~\cite{dias2020practical}:
	\begin{equation}\label{decompositionF}
		n_i=\frac{1}{h}\sum_R \chi(R)\chi_i(R),
	\end{equation}
	where $n_i$ is the number of times an IR, $i$, contributes to the reducible representation. $h$ is the order of the group, $\chi(R)$ is the character of the reducible representation for the operator $R$, and $\chi_i(R)$ is the character of the IR for the operator $R$. Each MO gives two spin orbitals of the same spatial symmetry. Therefore, there are six $A_1$ spin orbitals and two $B_1$ spin orbitals.
	
	We label the three $A_1$ MOs as $\mu_1, \mu_2, \mu_3$ and the $B_1$ MO as $\mu_4$. The bitstrings and occupied MOs in Table~\ref{tab: Fock_state} depend on this labeling. We order the eight qubits as
	\begin{equation}\label{bitString}
		|\mu_4(\downarrow),\mu_4(\uparrow),\mu_3(\downarrow),\mu_3(\uparrow),\mu_2(\downarrow),\mu_2(\uparrow),\mu_1(\downarrow),\mu_1(\uparrow)\rangle,
	\end{equation}
	where $\mu_i(s)$ represents the $i^{\text{th}}$ MO with spin $s$, in the left column of Table~\ref{tab: Fock_state}. Each qubit represents each spin orbital occupancy under the Jordan-Wigner transformation. The occupied MOs labeled by $(i,j,k)$ that hold spin-up, spin-up, and spin-down electrons respectively for $S_z=\frac{1}{2}$ are listed in the middle column of Table~\ref{tab: Fock_state}. The rightmost column in Table~\ref{tab: Fock_state} lists the decimal representations when the bitstrings are interpreted as binary numbers with the leftmost bit as the most significant bit. These decimal labels are just shorthand to show the corresponding Fock states on the $x$-axis of Figs.~\ref{fig:P_combo_2b}--\ref{fig:Counts_QPU}.
	
	Because the model Hamiltonian commutes with $N$, $S_z$, and all $C_{2v}$ operations, the sector $(N=3, S_z=+\tfrac{1}{2}, A_1)$ is invariant under the evolution. The initial Hartree-Fock determinant $|00000111\rangle$ belongs to this sector, so individual Fock states of the same symmetry are allowed in the adiabatically evolved ground state. Of the $\binom{8}{3}=56$ three-electron Fock states in the eight spin-orbital basis, the $S_z=\frac{1}{2}$ sector contains $\binom{4}{2}\binom{4}{1}=24$. A Fock state has $A_1$ spatial symmetry only if the direct product of the irreducible representations of its occupied spin orbitals is $A_1$. The three-electron configurations therefore must satisfy
	\begin{equation}\label{IR_product}
		O_i\otimes O_j\otimes O_k=A_1,
	\end{equation}
	where $O_i$ is the IR of the $i^{\text{th}}$ occupied spin orbital. Since $A_1$ is the identity representation and $B_1\otimes B_1=A_1$ based on the multiplication table [Table~\ref{tab: MultiplicationTable}], an $A_1$ spatial determinant must contain an even number of $B_1$ spin orbitals. With three electrons, this means the $B_1$ MO (i.e., $\mu_4$) is either empty or doubly occupied.
	
	If $\mu_4$ is empty, the three electrons are distributed among the three $A_1$ MOs. There are $\binom{3}{2}=3$ ways to choose the two $A_1$ MOs occupied by spin-up electrons and 3 ways to choose the $A_1$ MO occupied by the spin-down electron, giving 9 Fock states (i.e., the first 9 Fock states from the top of Table~\ref{tab: Fock_state}). If $\mu_4$ is doubly occupied, it must contain one spin-up and one spin-down electron, and the remaining spin-up electron occupies one of the three $A_1$ MOs, giving 3 additional Fock states (i.e., the last 3 Fock states in Table~\ref{tab: Fock_state}). Thus, there are $9+3=12$ Fock states with $A_1$ spatial symmetry and $S_z=\frac{1}{2}$. These are precisely the states listed in Table~\ref{tab: Fock_state}. The ground-state distributions of Figs.~\ref{fig:P_combo_2b}--\ref{fig:Counts_QPU} show nonzero weight only for these twelve states, in agreement with the symmetry requirement.
	
	\begin{table*}[h]
		\centering
		\caption{Fock states of $A_1$ symmetry and $S_z=\frac{1}{2}$ for 3 electrons in 8 spin orbitals}
		\label{tab: Fock_state}
		\begin{tabular}{ |c|c|c| }
			\hline
			Fock-state configuration & occupied MOs = $|i(\uparrow), j(\uparrow), k(\downarrow)\rangle$ & decimal label \\
			\hline
			|00000111$\rangle$ & |1,2,1$\rangle$ & 7 \\
			\hline
			|00001101$\rangle$ & |1,2,2$\rangle$ & 13 \\
			\hline
			|00010011$\rangle$ & |1,3,1$\rangle$ & 19 \\
			\hline
			|00010110$\rangle$ & |2,3,1$\rangle$ & 22 \\
			\hline
			|00011001$\rangle$ & |1,3,2$\rangle$ & 25 \\
			\hline
			|00011100$\rangle$ & |2,3,2$\rangle$ & 28 \\
			\hline
			|00100101$\rangle$ & |1,2,3$\rangle$ & 37 \\
			\hline
			|00110001$\rangle$ & |1,3,3$\rangle$ & 49 \\
			\hline
			|00110100$\rangle$ & |2,3,3$\rangle$ & 52 \\
			\hline
			|11000001$\rangle$ & |1,4,4$\rangle$ & 193 \\
			\hline
			|11000100$\rangle$ & |2,4,4$\rangle$ & 196 \\
			\hline
			|11010000$\rangle$ & |3,4,4$\rangle$ & 208 \\
			\hline
		\end{tabular}
	\end{table*}
	
	\setlength{\tabcolsep}{15pt} 
	\begin{table*}[h]
		\centering
		\caption{Transformation of Atomic Orbitals under $C_{2v}$}
		\label{tab: transAO}
		\begin{tabular}{|c|c|c|c|c|c|}
			\hline
			\multicolumn{2}{|c|}{} & \multicolumn{4}{c|}{\textbf{$C_{2v}$ symmetry operators}}\\\cline{3-6}
			\multicolumn{2}{|c|}{} & $E$ & $C_2$ & $\sigma_v(xy)$ & $\sigma_v(yz)$ \\\hline
			\multirow{4}{*}{\textbf{Hydrogen Atom}}& $\ce{H_a}$ & $\ce{H_a}$ & $\ce{H_a}$ & $\ce{H_a}$ & $\ce{H_a}$ \\\cline{2-6}
			& $\ce{H_b}$ & $\ce{H_b}$ & $\ce{H_b}$ & $\ce{H_b}$ & $\ce{H_b}$ \\\cline{2-6}
			& $\ce{H_c}$ & $\ce{H_c}$ & $\ce{H_d}$ & $\ce{H_c}$ & $\ce{H_d}$ \\\cline{2-6}
			& $\ce{H_d}$ & $\ce{H_d}$ & $\ce{H_c}$ & $\ce{H_d}$ & $\ce{H_c}$\\\hline
		\end{tabular}
	\end{table*}
	
	\begin{table*}[h]
		\centering
		\caption{Character Table of $C_{2v}$}
		\label{tab: CharacterTable}
		\begin{tabular}{|c|c|c|c|c|c|}
			\hline
			\multicolumn{2}{|c|}{} & \multicolumn{4}{c|}{\textbf{$C_{2v}$ symmetry operators}}\\\cline{3-6}
			\multicolumn{2}{|c|}{} & $E$ & $C_2$ & $\sigma_v(xy)$ & $\sigma_v(yz)$ \\\hline
			\multirow{4}{*}{\textbf{Irreducible Representation(IR)}}& $A_1$ & $1$ & $1$ & $1$ & $1$ \\\cline{2-6}
			& $A_2$ & $1$ & $1$ & $-1$ & $-1$ \\\cline{2-6}
			& $B_1$ & $1$ & $-1$ & $1$ & $-1$ \\\cline{2-6}
			& $B_2$ & $1$ & $-1$ & $-1$ & $1$\\\hline
		\end{tabular}
	\end{table*}
	
	\begin{table*}[h]
		\centering
		\caption{Multiplication Table of $C_{2v}$}
		\label{tab: MultiplicationTable}
		\begin{tabular}{|c|c|c|c|c|}
			\hline
			$\otimes$ & $A_1$ & $A_2$ & $B_1$ & $B_2$ \\\hline
			$A_1$ & $A_1$ & $A_2$ & $B_1$ & $B_2$ \\\hline
			$A_2$ & $\cdot$ & $A_1$ & $B_2$ & $B_1$ \\\hline
			$B_1$ & $\cdot$ & $\cdot$ & $A_1$ & $A_2$ \\\hline
			$B_2$ & $\cdot$ & $\cdot$ & $\cdot$ & $A_1$\\\hline
		\end{tabular}
	\end{table*}
\end{document}